\documentclass[american,aps,prb,reprint]{revtex4-1}
\usepackage{mathptmx}
\usepackage[LGR,T1]{fontenc}
\usepackage[utf8]{inputenc}
\usepackage{color}
\usepackage{babel}
\usepackage{textcomp}
\usepackage{amstext}
\usepackage{amssymb}
\usepackage{graphicx}
\usepackage[unicode=true,pdfusetitle,
 bookmarks=true,bookmarksnumbered=false,bookmarksopen=false,
 breaklinks=true,pdfborder={0 0 0},pdfborderstyle={},backref=false,colorlinks=true]
 {hyperref}
\hypersetup{
 allcolors=blue}

\makeatletter

\ProvideTextCommand{\~}{LGR}[1]{\char126#1}

\providecommand{\tabularnewline}{\\}

\bibpunct{[}{]}{,}{n}{}{}

\makeatother

\begin{document}
\title{Formation and dissociation reactions of complexes involving interstitial
carbon and oxygen defects in silicon}
\author{H. M. Ayedh}
\email{hussein.ayedh@smn.uio.no}

\address{University of Oslo, Department of Physics/Center for Materials Science
and Nanotechnology, P.O. Box 1048 Blindern, N-0316 Oslo, Norway}
\author{E. V. Monakhov}
\address{University of Oslo, Department of Physics/Center for Materials Science
and Nanotechnology, P.O. Box 1048 Blindern, N-0316 Oslo, Norway}
\author{J. Coutinho}
\address{I3N, Department of Physics, University of Aveiro, Campus Santiago,
3810-193 Aveiro, Portugal}
\begin{abstract}
We present a detailed first-principles study which explores the configurational
space along the relevant reactions and migration paths involving the
formation and dissociation of interstitial carbon-oxygen complexes,
$\mathrm{C_{i}O_{i}}$ and $\mathrm{C_{i}O_{2i}}$, in silicon. The
formation/dissociation mechanisms of $\mathrm{C_{i}O_{i}}$ and $\mathrm{C_{i}O_{2i}}$
are found as occurring via capture/emission of mobile $\mathrm{C_{i}}$
impurities by/from O-complexes anchored to the lattice. The lowest
activation energies for dissociation of $\mathrm{C_{i}O_{i}}$ and
$\mathrm{C_{i}O_{2i}}$ into smaller moieties are 2.3 eV and 3.1 eV,
respectively. The first is compatible with the observed annealing
temperature of $\mathrm{C_{i}O_{i}}$ , which occurs at around $400~^{\circ}$C,
and below the threshold for $\mathrm{O_{i}}$ diffusion. The latter
exceeds significantly the measured activation energy for the annealing
of $\mathrm{C_{i}O_{2i}}$ ($E_{\textnormal{a}}=2.55$~eV). We propose
that instead of dissociation, the actual annealing mechanism involves
the capture of interstitial oxygen by $\mathrm{C_{i}O_{2i}}$, thus
being governed by the migration barrier of $\mathrm{O_{i}}$ ($E_{\textnormal{m}}=2.53$~eV).
The study is also accompanied by measurements of hole capture cross
sections and capture barriers of $\mathrm{C_{i}O_{i}}$ and $\mathrm{C_{i}O_{2i}}$.
In combination with previously reported data, we find thermodynamic
donor transitions which are directly comparable to the first-principles
results. The two levels exhibit close features, conforming to a model
where the electronic character of $\mathrm{C_{i}O_{2i}}$ can be described
by that of $\mathrm{C_{i}O_{i}}$ perturbed by a nearby O atom. {[}\emph{Post-print
published in Physical Review Materials }\textbf{\emph{4}}\emph{, 064601
(2020)}; DOI:\href{https://doi.org/10.1103/PhysRevMaterials.4.064601}{10.1103/PhysRevMaterials.4.064601}{]}
\end{abstract}
\maketitle

\section{Introduction}

Understanding the evolution of point defect complexes in silicon (Si)
during device processing is highly significant to the industry of
electronics and photovoltaics. Carbon and oxygen are the most common
foreign species in mono-crystalline Si wafers grown by the Czochralski
technique (Cz-Si), with concentrations of about $10^{16}$~cm$^{-3}$
and over $10^{17}$~cm$^{-3}$, respectively. They are predominantly
present in the as grown Si material in the form of substitutional
carbon ($\mathrm{C_{s}}$) or interstitial oxygen ($\mathrm{O_{i}}$)
impurities.

Despite being electrically inert, $\mathrm{C_{s}}$ and $\mathrm{O_{i}}$
can have significant influence on the electrical properties of Si
wafers/samples. A conspicuous example is the formation of the so-called
thermal double donors, which comprise a family of oxygen aggregates
that lead to a considerable increase in free-electron concentration.
The effect occurs upon heat treatments of O-rich Si in the temperature
range of 400-500~$^{\circ}\mathrm{C}$, while it is strongly suppressed
in samples where in addition to oxygen, carbon is also present in
a high concentration \citep{Kaiser1958,Bean1972}. Another prominent
example is the so-called light-induced-degradation (LID) of $\textrm{n}^{+}\textrm{p}$-Si
solar cells where the p-type layer is boron (B) doped \citep{Schmidt2003,Schmidt2004}.
Recent measurements and theory indicate that the defect comprises
a complex made of one boron atom and an oxygen dimer \citep{Vaqueiro-Contreras2019}.

Both $\mathrm{C_{s}}$ and $\mathrm{O_{i}}$ can trap intrinsic defects,
resulting in electrically active complexes that can travel long distances
in the Si \citep{Christopoulos2016}. Interstitial oxygen comprises
a two-fold coordinated O atom sitting near the center of Si-Si bonds.
Migration of $\mathrm{O_{i}}$ proceeds via consecutive jumps between
nearest bonds with an activation energy $E_{\textnormal{m}}=2.53$~eV
\citep{Newman2000}. It is primarily a trap for vacancy-type defects,
in particular the mono-vacancy ($V$), thus leading to formation of
the vacancy-oxygen complex which has an acceptor level at $E_{\textnormal{c}}-0.17$~eV
\citep{Watkins1961,Lee1976}.

On the other hand, $\mathrm{C_{s}}$ is an efficient trap for self-interstitials
($\mathrm{Si_{i}}$), where $\mathrm{Si_{i}}$ partially takes the
place of the C-atom, which is displaced from the substitutional site
to become interstitial carbon ($\mathrm{C_{i}}$). This reaction has
been termed as \emph{kick-out mechanism} \citep{Watkins1965}. $\mathrm{C_{i}}$
shows a split-interstitial configuration, comprising a C-Si dimer
sharing a lattice site and aligned along $\langle001\rangle$ \citep{Watkins1976,Lee1977,Leary1997}.
$\mathrm{C_{i}}$ produces both acceptor ($E_{\mathrm{c}}-0.10$~eV)
and donor ($E_{\mathrm{v}}+0.28$~eV) levels \citep{Londos1987}
($E_{\mathrm{c}}$ and $E_{\mathrm{v}}$ denote the conduction and
valence band edge energies respectively). The defect becomes mobile
just above room-temperature (RT), and depending on doping type, anneals
out with an activation energy of $\sim0.7\textrm{-}0.87$~eV (assigned
to its migration barrier), yielding a diffusivity in the range 10-15~cm$^{2}$/s
\citep{Watkins1976,Tipping1987,Tersoff1990}

In carbon-rich Si, mobile $\mathrm{C_{i}}$ defects are most likely
to be captured by $\mathrm{C_{s}}$, forming interstitial-carbon-substitutional-carbon
complexes ($\mathrm{C_{i}C_{s}}$) \citep{ODonnell1983}. In oxygen-rich
Si, the dominant trap for the mobile $\mathrm{C_{i}}$ is $\mathrm{O_{i}}$
\citep{Londos1988}, leading to formation of interstitial-carbon-interstitial-oxygen
complexes ($\mathrm{C_{i}O_{i}}$). $\mathrm{C_{i}O_{i}}$ has a deep
donor level at $E_{\mathrm{v}}+0.36$~eV, and that has been corroborated
by both electron paramagnetic resonance (EPR) and deep-level transient
spectroscopy (DLTS) measurements \citep{Trombetta1987,Asom1987}.
We should add that substitutional boron ($\mathrm{B_{s}}$) is known
to compete with $\mathrm{C_{s}}$ for the capture of $\mathrm{Si_{i}}$
defects by a similar kick-out mechanism, suppressing the generation
of $\mathrm{C_{i}}$, and therefore, the formation of $\mathrm{C_{i}O_{i}}$
\citep{Mooney1977,Kimerling1989}. When $\mathrm{B_{s}}$ is the dominant
trap, the resulting boron interstitial ($\mathrm{B_{i}}$) defects
become mobile at RT and ultimately form $\mathrm{B_{i}O_{i}}$ and
$\mathrm{B_{i}B_{s}}$ complexes, depending on the relative content
of $\mathrm{B_{s}}$ and $\mathrm{O_{i}}$ in the samples \citep{Mooney1977,Kimerling1989,Watkins1975,Vines2008}.

$\mathrm{C_{i}O_{i}}$ has been extensively investigated by several
spectroscopic techniques. The defect gives rise to a conspicuous zero-phonon
emission line at 789 meV, the so-called C-line in the photoluminescence
(PL) spectra, at low temperatures ($T\leq20$~K) \citep{Thonke1984,Davies1986,Kuerner1989}.
This energy is close to the gap width ($E_{\textrm{g}}$) subtracted
by the hole binding energy of the donor level ($E_{\text{C-line}}\approx E_{\text{g}}-0.36$~eV),
and was interpreted as resulting from the recombination energy of
a diffuse electron possessing a conduction-band-like character with
a hole tightly bound ($\sim0.36$~eV binding energy) to neutral $\mathrm{C_{i}O_{i}}$.
From Fourier-transform infra-red (FTIR) absorption spectroscopy, two
main local vibrational mode (LVM) bands associated with $\mathrm{C_{i}O_{i}}$
are also well established. At low temperatures, they appear at 865~cm$^{-1}$
(known as C(3) band) and at 1116~cm$^{-1}$ \citep{Davies1986,Svensson1986a,Coutinho2001,Murin2001}.
From \emph{ab-initio} local density functional calculations, Jones
and Öberg \citep{Jones1992} and more recently Coutinho \emph{et~al.}
\citep{Coutinho2001}, showed that in the $\mathrm{C_{i}O_{i}}$ ground
state, both O and C are three-fold coordinated. This configuration
was shown to be necessary in order to account for the electrical,
optical and magnetic resonance experiments \citep{Trombetta1987,Davies1986,Svensson1986a,Coutinho2001,Murin2001}

More recently, Khirunenko \emph{et~al.} \citep{Khirunenko2005,Khirunenko2008}
found that the formation of $\mathrm{C_{i}O_{i}}$ is more complex
than previously thought, and reported the formation of metastable
configurations of $\mathrm{C_{i}O_{i}}$ during isochronal annealing
of irradiated Si in the temperature range 280-360~K. These findings
were interpreted as the formation of a precursor (labelled $\mathrm{C_{i}O_{i}^{*}}$)
before reaching the ground state \citep{Khirunenko2005,Khirunenko2008,Abdullin1990}.
The proposed geometry for the metastable defect consisted of $\mathrm{C_{i}}$
and $\mathrm{O_{i}}$ defects lying on a common $\{110\}$ plane and
separated by a Si-Si bond, essentially retaining their three-fold
and two-fold coordination of their individual structures, respectively.
From annealing data, the $\mathrm{C_{i}O_{i}^{*}}\rightarrow\mathrm{C_{i}O_{i}}$
conversion was estimated to be activated by a barrier of $\sim1$~eV
\citep{Abdullin1990}. Although capture/transformation/dissociation
mechanisms have not been explored by theory, the binding energy of
$\mathrm{C_{i}O_{i}^{*}}$ was calculated to be $\sim$0.7~eV, about
1~eV lower than that estimated for $\mathrm{C_{i}O_{i}}$ (ground
state) \citep{Khirunenko2008}.

The evolution of $\mathrm{C_{i}O_{i}}$ upon post-irradiation thermal
treatments has been studied for decades, but surprisingly, few studies
addressed the kinetics and annealing mechanisms of $\mathrm{C_{i}O_{i}}$.
The defect is generally considered to be stable up to 400 °C \citep{Kimerling1989,Kuerner1989,Svensson1986a}.
By monitoring the C(3) absorption band in high-fluence MeV electron-irradiated
Cz samples \citep{Svensson1986a}, first order annealing kinetics
was inferred with activation energy and pre-exponential factor of
$2.0$~eV and $3\times10^{12}$~s$^{-1}$, respectively. On the
basis of these results, $\mathrm{C_{i}O_{i}}$ was tentatively suggested
to anneal out via dissociation ($\mathrm{C_{i}O_{i}}\rightarrow\mathrm{C_{i}}+\mathrm{O_{i}}$).

From PL measurements, the loss of the C-line during heat treatments
at 350-450~°C has been observed to be accompanied by the formation
of the so-called P-line at 767~meV, \citep{Kuerner1989,Davies1994,Davies2006}.
Several photoluminescence studies have argued that the P-line is associated
with carbon- and oxygen-related defects, most likely involving an
oxygen dimer bonded to $\mathrm{C_{i}}$ \citep{Kuerner1989,Davies1994,Davies2006,Fukuoka1990,Awadelkarim1990}.
The PL spectra of the C- and P-lines exhibit almost identical properties
as well as very similar effective-mass like excited states. Furthermore,
like the C-center, the point group symmetry of the P-center is also
$C_{1h}$.

Assuming that like the C-line, the P-line results from recombination
of an effective-mass-like electron with a hole on a deep donor state,
from the difference of their zero-phonon energies we may infer that
the P-center gives rise to a donor transition 22~meV above that of
$\mathrm{C_{i}O_{i}}$, \emph{i.e.,} at about $E_{\text{v}}+0.38$~eV.
In fact, a correlation has been recently found between the P-line
and a DLTS peak at $E_{\textrm{v}}+0.39$~eV that forms upon the
annealing of $\mathrm{C_{i}O_{i}}$ \citep{Ganagona2012,Raeissi2013}.
This peak was tentatively assigned to an interstitial-carbon-interstitial-dioxygen
complex ($\mathrm{C_{i}O_{2i}}$). The annealing of $\mathrm{C_{i}O_{i}}$
was proposed to occur via dissociation into $\mathrm{C_{i}}$ and
$\mathrm{O_{i}}$, with the released $\mathrm{C_{i}}$ defects being
subsequently trapped by $\mathrm{O_{2i}}$ \citep{Ganagona2012}.
Considering that the above suggests that trapping of $\mathrm{C_{i}}$
by progressively larger oxygen aggregates tends to raise a donor level
towards higher energies within the band gap, it is important to determine
the mechanisms involved in the operating reactions, and ultimately
the impact of the reaction products in terms of recombination power.

Recently, Ayedh \emph{et~al.} \citep{Ayedh2019} studied the annealing
kinetics of $\mathrm{C_{i}O_{2i}}$ in irradiated p-type (B-doped)
Cz-Si samples by monitoring the $E_{\textrm{v}}+0.39$~eV hole trap
via DLTS. The trap anneals out according to first order kinetics,
exhibiting an activation energy of $\sim$2.55 eV, and a pre-exponential
factor in the range $(2\textrm{-}30)\times10^{12}$~s$^{-1}$. From
the kinetics and deduced pre-factor (of the order of the Debye frequency
of Si), it was suggested that the annealing of $\mathrm{C_{i}O_{2i}}$
occurs via dissociation, rather than being a diffusion-limited process.
To clarify this and other issues related to the formation and dissociation
of carbon-oxygen complexes, we performed a detailed theoretical study
which explores the configurational space along relevant reaction and
migration paths involving C and O species in Si. The study is also
accompanied by measurements of capture cross-sections and barriers
for the capture of holes by $\mathrm{C_{i}O_{i}}$ and $\mathrm{C_{i}O_{2i}}$,
allowing us to accurately locate their thermodynamic donor transition.

\section{Method details\label{sec:method}}

\subsection{Calculation details\label{subsec:method-theory}}

First-principles calculations were performed using the Vienna Ab-initio
Simulation Package (VASP) \citep{Kresse1993,Kresse1994,Kresse1996,Kresse1996a},
employing the projector-augmented wave (PAW) method for the treatment
of the core electrons \citep{Blochl1994}. A basis set of plane-waves
with kinetic energy of up to 400~eV was used to describe the Kohn-Sham
states. All many-body energies reported were evaluated self-consistently,
using the hybrid density functional of Heyd-Scuseria-Ernzerhof (HSE06)
\citep{Heyd2003,Krukau2006}, up to a numerical accuracy of $10^{-7}$~eV.
Hybrid density functionals perform relatively better in terms of the
calculated band structure, when compared with semi-local functionals,
including those using the generalized gradient approximation (GGA)
\citep{Perdew1996}. The GGA was however employed for the search of
ground-state and saddle-point structures along the minimum energy
paths (MEP) of atomistic mechanisms. All structures were optimized
until forces on atoms were below 0.01~eV/Å. The two-step method ---
firstly involving a calculation of a GGA-level structure followed
by a single-point energy calculation within HSE06, was shown to lead
to numerical error bars below $10$~meV for relative energies, including
formation energies and migration barriers \citep{Gouveia2019,Bathen2019}.
These tests also apply to the methods used below to find saddle point
energies, which are simply a sequence of structural relaxations subject
to a particular set of constraints.

We used 216-atom supercells of silicon (with cubic shape), obtained
by replication of $3\!\times\!3\!\times\!3$ conventional cells, in
which carbon and oxygen atoms were inserted to produce CO-related
defects. The defects considered were all interstitials, namely carbon
($\mathrm{C_{i}}$), oxygen ($\mathrm{O_{i}}$), oxygen dimer ($\mathrm{O_{2i}}$),
carbon-oxygen ($\mathrm{C_{i}O_{i}}$) and carbon-dioxygen ($\mathrm{C_{i}O_{2i}}$).
The equilibrium (calculated) lattice parameters of Si was $a=5.4318$~Å,
matching the experimental value of $a=5.4310$~Å. The Brillouin zone
(BZ) of GGA- and HSE06-level calculations was sampled at $\Gamma$centered
$2\!\times\!2\!\times\!2$ ($\Gamma$-$2^{3}$) and $1\!\times\!1\!\times\!1$
($\Gamma$-point) $\mathbf{k}$-point meshes, respectively.

To investigate defect migration and transformation processes, we employed
a combination of \emph{nudged elastic band} (NEB) \citep{Mills1994,Mills1995}
and \emph{dimer} \citep{Henkelman1999} methods (at the GGA level).
For the NEB calculations, initial and final (frozen) geometries were
at the limits of a sequence of 9-11 intermediate images, which were
created at first hand by linear interpolation and adjusted to avoid
unphysical bond lengths. Saddle-point search calculations involved
a first step consisting of a fast exploratory NEB run with the Brillouin
zone being sampled at the $\Gamma$-point. On a subsequent step, we
increased the $\mathbf{k}$-point sampling density to $\Gamma$-$2^{3}$,
and refined the exploratory MEP by either employing the climbing-image
NEB method \citep{Henkelman2000} or by performing a dimer search.
The dimer run was initiated using the two higher-energy structures
obtained from the previous exploratory NEB step. Finally, the resulting
highest-energy configuration along each MEP (the saddle point) was
taken in order to calculate its total energy within HSE06.

\subsection{Measurement details}

A set of $\mathrm{n^{+}p}$ diodes were prepared (fabrication details
are described in Ref.~\citep{Ayedh2019}) on a p-type (boron doped)
Cz-Si wafer with resistivity of $\sim14$~$\Omega$\,cm, corresponding
to a net carrier concentration of about $\thicksim1\times10^{15}$~cm$^{-3}$
at RT, as determined by capacitance-voltage (C-V) measurements with
a $1\mathrm{MHz}$ probe frequency. The oxygen and carbon concentration
in the wafers was $7\times10^{17}$~cm$^{-3}$ and $\leq2\times10^{16}$~cm$^{-3}$,
respectively, as determined by secondary ion mass spectrometry (SIMS).
Aluminum (Al) Ohmic contacts were deposited by electron beam evaporation
on the front side ($\mathrm{n^{+}}$ layer) and silver paste was applied
on the back side of the samples to form an Ohmic contact. The fabricated
$\mathrm{n^{+}p}$ diodes were subjected to annealing at $300$~℃
for $26$~h in $\mathrm{N_{2}}$ atmosphere and then were irradiated
with $1.8\:\mathrm{MeV}$ protons at RT to doses of $1\times10^{13}$~cm$^{-2}$.
After irradiation, the samples were annealed at $400$~℃ for $1.5$~h
in order to anneal out all the radiation-produced point defects, except
$\mathrm{C_{i}O_{i}}$ which exhibited a dominating DLTS peak. One
of the samples was subjected to a multiple-step annealing at $400$~℃
for about 30~h in total in order to achieve a complete annealing
of the $\mathrm{C_{i}O_{i}}$ complex and formation of $\mathrm{C_{i}O_{2i}}$
(see Ref.~\citep{Ayedh2019} for further details). C-V and DLTS measurements
were employed after each annealing step for characterizing the samples
using a refined version of the setup described in Ref.~\citep{Svensson1989},
equipped with a closed-cycle He cryostat.

In DLTS, the reverse bias quiescent voltage was kept at $-10$~V,
the filling pulse was 50~ms long at 0~V bias, and the sample temperature
was scanned between 50~K and 300~K. The DLTS signal was extracted
from the recorded capacitance transients applying a lock-in and a
high resolution weighting function, so-called GS4 \citep{Istratov1997},
with six different rate windows in the range of $(20\textnormal{-}640\,\textnormal{ms})^{-1}$.
The $\mathrm{C_{i}O_{i}}$ and $\mathrm{C_{i}O_{2i}}$ defects were
monitored via their respective deep levels at $\mathrm{E_{\textrm{v}}+0.36}$~eV
and $\mathrm{E_{\textrm{v}}+0.39}$~eV. These correspond to DLTS
peak positions at 173~K and 190~K, respectively, when employing
a rate window of 640~ms$^{-1}$ and the GS4 weighting functions.
Hole capture cross sections ($\sigma$) were measured for $\mathrm{C_{i}O_{i}}$
and $\mathrm{C_{i}O_{2i}}$ traps by varying the filling pulse duration
from 10~ns to 10~$\mathrm{\mu s}$ and recording the amplitude of
the level signal. During these measurements a single rate window was
used for each capture cross section measurement and the sample temperature
was kept constant within $T=T_{\textnormal{max}}\pm0.1\:$~K, where
$T_{\textnormal{max}}$ is the temperature that corresponds to to
maximum DLTS signal for a specific rate window.

\section{Results\label{sec:results}}

\subsection{Migration energies of elementary defects}
\noindent \begin{flushleft}
\begin{figure}
\begin{centering}
\includegraphics[width=7cm]{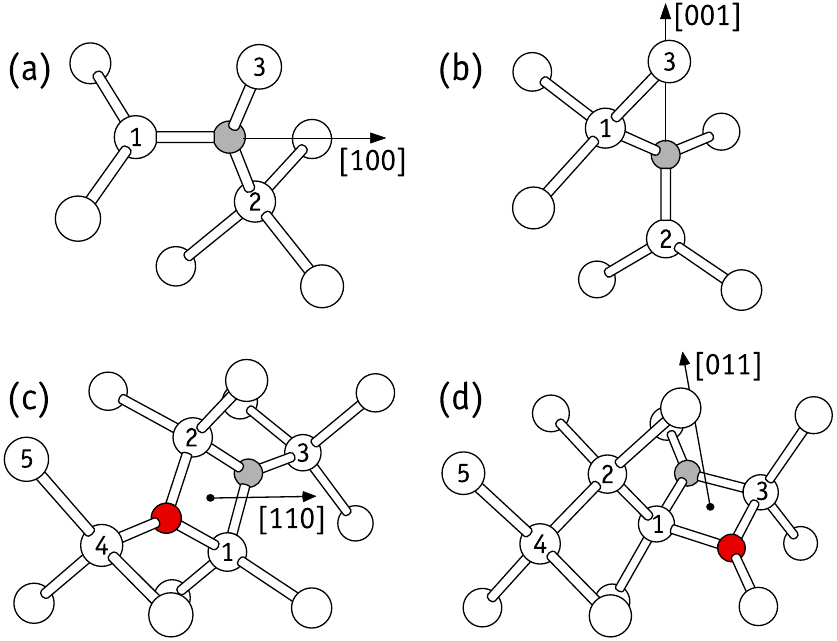}
\par\end{centering}
\caption{\label{fig1}Initial (a,c) and final (b,d) structures for single jumps
during the migration of $\mathrm{C_{i}}$ and $\mathrm{C_{i}O_{i}}$
defects in silicon. Some ligands are labeled with numbers in order
to assist the reader in the identification of the Si neighbours before
and after the jumps. Oxygen, carbon and silicon atoms are represented
in red, gray and white, respectively. Some crystallographic directions
are also represented to indicate the alignment of the defects with
respect to the host lattice.}
\end{figure}
\par\end{flushleft}

We start by reporting on some properties of basic elements that participate
in the reactions addressed in Sec.~\ref{sec:reactions}, namely $\mathrm{O_{i}}$,
$\mathrm{O_{2i}}$, $\mathrm{C_{i}}$, $\mathrm{C_{i}}\mathrm{O_{i}}$
and $\mathrm{C_{i}}\mathrm{O_{2i}}$.

In agreement with previously experimental and theoretical studies
\citep{Watkins1976,Lee1977,Leary1997,Song1990}, we find that the
ground-state structure of interstitial carbon ($\mathrm{C_{i}}$)
in Si comprises a C-Si split-interstitial, possessing dangling bonds
on both C and Si atoms. As shown in Figs.~\ref{fig1}(a) and \ref{fig1}(b),
the C-Si dimer is aligned along the $\langle001\rangle$ direction,
and both atoms share a lattice site.

As for interstitial oxygen $(\mathrm{O_{i})}$ in Si, the resulting
ground state structure is also in line with the widely accepted model,
according to which the O atom sits near the bond center site, forming
a \emph{puckered} Si-O-Si unit \citep{Bosomworth1970,Pesola1999,Coutinho2000}.
The orbiting motion of the O atom around the $\langle111\rangle$
axis of the perfect Si-Si bond, as well as its motion across the bond
center site, involve surmounting rather shallow energy barriers of
the order of 10~meV.

The minimum-energy structure of the oxygen dimer is the so-called
\emph{staggered} configuration \citep{Pesola1999,Coutinho2000,Needels1991},
where two O atoms occupy neighboring bond center sites, thus connecting
to a common Si atom. In this geometry, the O atoms in the Si-O-Si-O-Si
structure are displaced in a staggered way. This minimizes the departure
from the sp$^{3}$ bond angles involving the central Si atom and its
O neighbors \citep{Pesola1999,Coutinho2000,Needels1991}.

For the $\mathrm{C_{i}O_{i}}$ and $\mathrm{C_{i}O_{2i}}$ complexes,
we found that in both cases the defect core comprises a square-like
structure involving C, O and two Si atoms. Figs.~\ref{fig1}(c) and
\ref{fig1}(d) represent two orientations of $\mathrm{C_{i}O_{i}}$
in Si. For the case of $\mathrm{C_{i}O_{2i}}$, the structure is similar
to that of $\mathrm{C_{i}O_{i}}$, although an additional O atom is
connected to Si atoms number 4 and 5 in Figs.~\ref{fig1}(c). Hence,
the complex is essentially a staggered $\mathrm{O_{2i}}$ next to
$\mathrm{C_{i}}$.

We also found a metastable $\mathrm{C}_{\mathrm{i}}\mathrm{O}_{\mathrm{i}}^{*}$
complex, identical to that reported by Khirunennko \emph{et~al.}
\citep{Khirunenko2008}, consisting of C$_{\textnormal{i}}$ and two-fold
coordinated O$_{\textnormal{i}}$ impurities separated by a Si-Si
bond and sharing the same plane. The $\mathrm{C}_{\mathrm{i}}\mathrm{O}_{\mathrm{i}}^{*}$
complex was found 1.0~eV above the ground state, \emph{i.e.}, with
a binding energy of 0.56~eV with respect to isolated C$_{\textnormal{i}}$
and O$_{\textnormal{i}}$ impurities.

In the analysis of the defect reactions described in Sec.~\ref{sec:reactions},
we assume that the migration barriers for C$_{\textnormal{i}}$, O$_{\textnormal{i}}$
and O$_{\textnormal{2i}}$ are respectively $(0.73\pm0.05)$~eV,
$(2.53\pm0.03)$~eV and $(2.02\pm0.01)$~eV as derived from experiments
\citep{Lastovskii2017,Corbett1964,Quemener2015}. The migration mechanisms
of all three defects are well established theoretically \citep{Capaz1994,Coutinho2000}.
Hence, a calculation of the respective barriers would provide us with
an idea of the error bar of the methodology. As mentioned in Sec.~\ref{subsec:method-theory},
the search for saddle-points was conducted along the configurational
space between initial and final ground states structures.

The MEP for migration of $\mathrm{C_{i}}$ was found to involve a
change in the defect alignment within the Si lattice. Initial and
final configurations are illustrated in Figs.~\ref{fig1}(a) and
\ref{fig1}(b). Accordingly, the C atom performs an \emph{out-of-plane}
jump, leading to a change of orientation of the main $\langle100\rangle$
symmetry axis. This mechanism was earlier found by Capaz \emph{et~al.}
\citep{Capaz1994}, and according to our calculations corresponds
to a MEP with a total barrier of $E_{\textnormal{m}}=0.76$~eV. This
figure matches recent annealing experiments carried out in $\textnormal{n}^{+}\textnormal{p}$-diodes
under reverse bias \citep{Lastovskii2017}, where a barrier for the
migration of carbon interstitial in the neutral charge state was measured
as $E_{\textnormal{m}}=0.73$~eV. The energy barrier for the \emph{in-plane}
jump of $\mathrm{C_{i}}$ is about 1.6~eV high, essentially due to
the fact that the carbon atom has to travel through a high-energy
two-fold coordinated Si-C-Si structure.

Interstitial oxygen migrates by hopping between neighboring (puckered)
bond center sites. The saddle-point configuration is commonly referred
to as Y-lid \citep{Coutinho2000}, and it is analogous to the stable
configuration of $\mathrm{C_{i}}$ depicted in Figure~\ref{fig1}
(the C atom being replaced by O). Our calculations indicate that the
Y-lid structure of $\mathrm{O_{i}}$ is 2.63~eV above the ground
state, overestimating the experimental figure by a mere 0.1~eV. This
result is in line with the 2.7~eV barrier height reported previously
using hybrid density functional calculations \citep{Binder2014}.
As for the oxygen dimer, the saddle-point for migration is attained
when both O atoms display three-fold coordination (see Figure~7 on
Ref~\citep{Coutinho2000}). Here we found that the relevant structure
along the path is 1.87~eV above the ground state, about 0.1~eV below
the measured value \citep{Quemener2015}. These results suggest that
the error bar regarding the energy barriers to be discussed below
is about 0.1~eV. This is approximately twice the error in the measured
barrier of $\mathrm{C_{i}}$ and about 10 times the error in the measured
barriers of $\mathrm{O_{i}}$ and $\mathrm{O_{2i}}$ \citep{Lastovskii2017,Corbett1964,Quemener2015}.

Like the carbon interstitial, the MEP for migration $\mathrm{C_{i}O_{i}}$
also involves a change in the defect alignment within the Si lattice.
Initial and final configurations are shown in Figs.~\ref{fig1}(c)
and \ref{fig1}(d). C and O atoms in $\mathrm{C_{i}O_{i}}$ jump off
the $\{110\}$ symmetry plane, in a sequential manner --- first the
carbon atom, then the oxygen. The second step has the highest barrier,
leading to an overall migration barrier of $E_{\textnormal{m}}=2.45$~eV.
This is 0.65~eV lower than the barrier for in-plane migration, and
such a large figure derives from the nearly independent jump of the
O atom.

Below we will argue that migration of $\mathrm{C_{i}O_{i}}$ is actually
an unlikely event due to the fact that dissociation is governed by
a lower barrier. An analogous argument applies to $\mathrm{C_{i}O_{2i}}$.

\subsection{Formation and dissociation of C$_{\text{i}}$O$_{\text{i}}$\textmd{\normalsize{}
}and C$_{\text{i}}$O$_{\text{2i}}$\label{sec:reactions}}

Before proceeding, we leave a few words about notation. Let us consider
that A and B stand for either $\textrm{O}_{\textrm{i}}$, $\textrm{C}_{\textrm{i}}$
or a complex made of any (including more than one) of these species.
Infinitely separated defects A and B are represented as $\textrm{A}+\textrm{B}$
(\emph{e.g.} $\mathrm{C_{i}O_{i}}+\mathrm{O_{i}}$); Complexes involving
close pairs of A and B atoms separated by two or more Si atoms, but
still sharing the same supercell volume, are represented as A-B (\emph{e.g.}
$\mathrm{C_{i}O_{i}}\text{-}\mathrm{O_{i}}$); Complexes involving
A and B moieties connected to a common Si atom are termed AB (\emph{e.g.}
$\mathrm{C_{i}O_{2i}}$). We did not find stable complexes involving
direct C-O or O-O bonds. Di-carbon complexes were not investigated.

Regarding the designation of the saddle-point structure along a particular
MEP, we found it useful to highlight the atoms that move most during
the respective mechanism. This is done by enclosing the moving species
within parentheses. For the dissociation of $\mathrm{C_{i}O_{2i}}$,
for instance, if the jumping moiety is the carbon atom (leaving $\mathrm{O_{2i}}$
behind), the carbon species is enclosed within parentheses. Hence,
a first-step for the dissociation reaction is cast as,

\begin{equation}
\mathrm{C_{i}O_{2i}}\stackrel{\mathrm{(C_{i})O_{2i}}}{\longrightarrow}\mathrm{C_{i}\textrm{-}O_{2i}},\label{cr:notation1}
\end{equation}
which involves a detachment of the carbon atom from $\mathrm{O_{2i}}$,
followed by a second step

\begin{equation}
\mathrm{C_{i}\textrm{-}O_{2i}}\stackrel{\mathrm{(C_{i})+O_{2i}}}{\longrightarrow}\mathrm{C_{i}+O_{2i}},\label{cr:notation2}
\end{equation}
describing the migration of $\mathrm{C_{i}}$ away from $\mathrm{O_{2i}}$.
\noindent \begin{flushleft}
\begin{figure*}
\begin{centering}
\includegraphics[width=16cm]{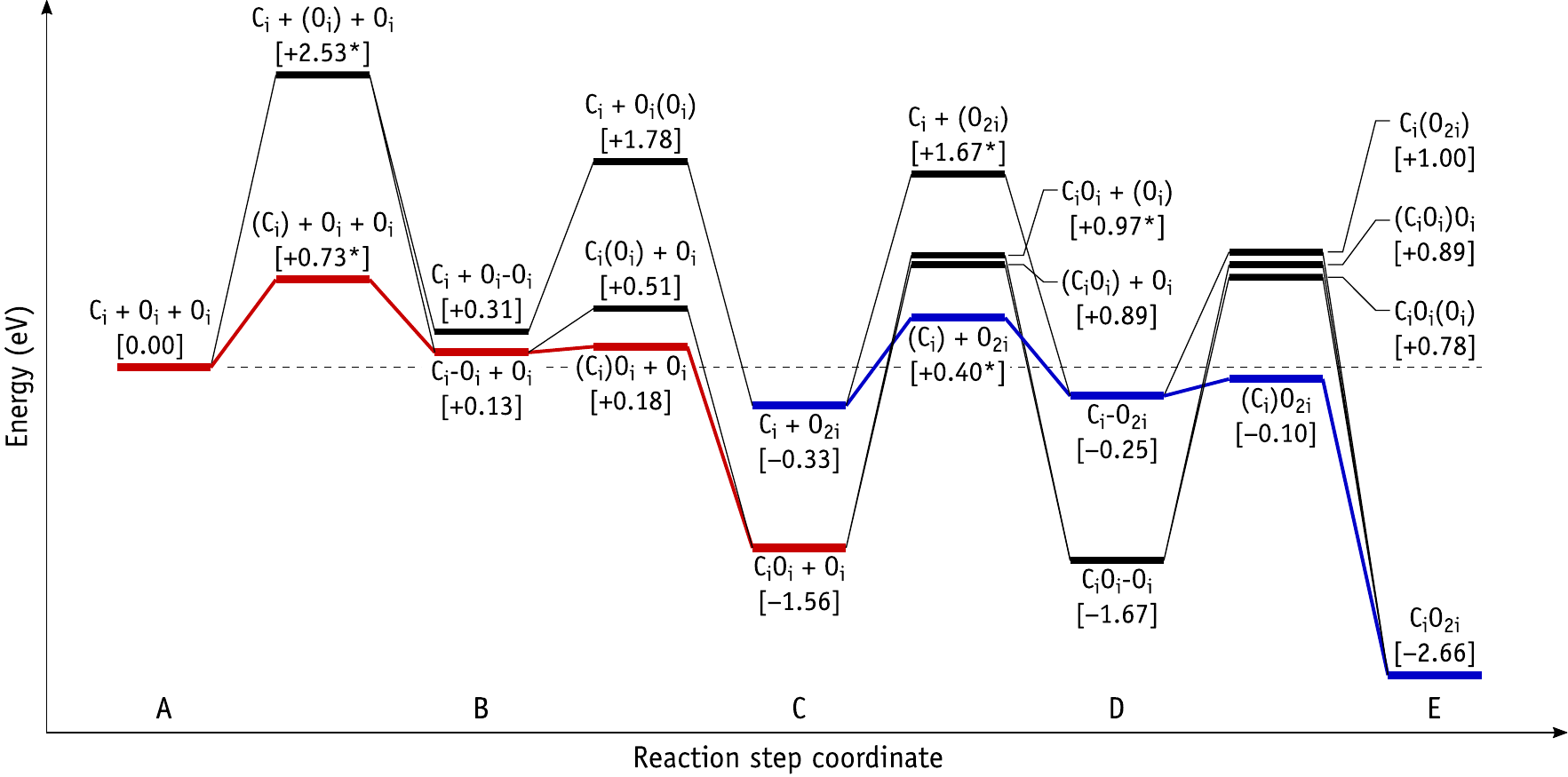}
\par\end{centering}
\caption{\label{fig2}Reaction energy diagram involving interstitial carbon
($\text{C}_{\text{i}}$) and two interstitial oxygen impurities ($\text{O}_{\text{i}}$)
in silicon. The diagram highlights in red and blue the most favorable
reactions paths describing the formation/dissociation of $\mathrm{\mathrm{C_{i}O_{i}}}$
and $\mathrm{\mathrm{C_{i}O_{2i}}}$ complexes, respectively. Reactions
proceed through sequential transformations between reaction steps
denoted with letters `A-E'. These correspond to stable states represented
by a thick horizontal segment, a state label and its relative energy
within square brackets. See text for details regarding label notation.
Intermediate states (between stable steps) are saddle-points and correspond
to the summit of the minimum energy path that separates neighboring
stable states. Atoms or groups of atoms that jump during each step
are represented within parentheses. The origin of the energy coordinate
is shown as a dashed horizontal line. All energies are in eV.}
\end{figure*}
\par\end{flushleft}

Figure~\ref{fig2} shows a reaction energy diagram accounting for
the interactions between one C and two O atoms in silicon. Each state
involves three interstitial atoms, being represented by a horizontal
segment in the energy scale, a state label and its relative energy,
in eV, enclosed within square brackets. Starred energies were obtained
with help of experimentally obtained migration barriers of $\mathrm{C_{i}}$,
$\mathrm{O_{i}}$ and $\mathrm{O_{2i}}$. All energies are relative
to the $\mathrm{C_{i}}+\mathrm{O_{i}}+\mathrm{O_{i}}$ state in step~A.
This is indicated by the wide horizontal dashed line, and stands for
uncorrelated $\text{C}_{\text{i}}$ and two $\text{O}_{\text{i}}$
impurities. Each step (A-E) groups one or more stable (ground or metastable)
states. Different states within each step are not necessarily close
in configurational space. States between neighboring steps are saddle-points.
The purpose of the thin lines connecting the horizontal state segments
is to relate every pair of stable states with at least one saddle
point. They also provide guidance to the reader in the identification
of mechanisms along the reactions steps. The atoms that move most
during a transition are enclosed within parentheses (see above).

Figure~\ref{fig2} also highlights the most favorable reactions paths
for the formation/dissociation of $\mathrm{\mathrm{C_{i}O_{i}}}$
($\textnormal{A}\leftrightarrow\textnormal{C}$) and $\mathrm{\mathrm{C_{i}O_{2i}}}$
($\textnormal{C}\leftrightarrow\textnormal{E}$) complexes in red
and blue, respectively. Formation and dissociation processes read
from left to right and vice-versa, respectively. The ground states
in steps C and E correspond to the $\mathrm{C_{i}O_{i}}$ and $\mathrm{C_{i}O_{2i}}$
complexes, respectively. From their relative energies we find that
the binding energies of $\mathrm{C_{i}}$ to $\mathrm{O_{i}}$ and
to $\mathrm{O_{2i}}$ are $E_{\textnormal{b}}=1.56$~eV and 2.33~eV
as obtained from the energy balance of,

\begin{equation}
\mathrm{C_{i}}+\mathrm{O_{i}}+\mathrm{O_{i}}\;(\textnormal{A})\rightarrow\mathrm{C_{i}O_{i}}+\mathrm{O_{i}}\;(\textnormal{C})+[E_{\textnormal{b}}\!=\!1.56\,\textnormal{eV}]\label{cr:eb-cioi}
\end{equation}
and

\begin{equation}
\mathrm{C_{i}}+\mathrm{O_{2i}}\;(\textnormal{C})\rightarrow\mathrm{C_{i}O_{2i}}\;(\textnormal{E})+[E_{\textnormal{b}}\!=\!2.33\,\textnormal{eV}],\label{cr:eb-cio2i}
\end{equation}
respectively, where reaction steps are indicated within parentheses.
The magnitude of the binding energies indicate a strong thermodynamic
drive for formation of these complexes. The binding energy of $\mathrm{C_{i}O_{i}}$
is also in line with previous calculations, where $E_{\textnormal{b}}$
values were found in the range 1.6-1.7~eV \citep{Coutinho2001,Khirunenko2008,Backlund2007}.
The binding energy between two interstitial O atoms is found from
the energy difference of states in steps A and C, namely

\begin{equation}
\mathrm{C_{i}}+\mathrm{O_{i}}+\mathrm{O_{i}}\;(\textnormal{A})\rightarrow\mathrm{C_{i}}+\mathrm{O_{2i}}\;(\textnormal{C})+E_{\textnormal{b}},
\end{equation}
with $E_{\textnormal{b}}=0.33$~eV, nicely matching the 0.3~eV found
experimentally by Murin et~al. \citep{Murin1998}.

Migration barriers of $\mathrm{C_{i}}$ (0.73~eV), $\mathrm{O_{i}}$
(2.53~eV), $\mathrm{O_{2i}}$ (2.02~eV) and $\mathrm{C_{i}}\mathrm{O_{i}}$
(2.45~eV) were considered for the saddle-point transitions $\textnormal{A}\rightarrow\textnormal{B}$
and $\textnormal{C}\rightarrow\textnormal{D}$. The reactants are
at least two independent defects, one of which is a diffusing species
(indicated within parentheses). The products involve metastable precursors
(steps B and D) that can be converted into stable defects in steps
C and E, respectively. The mechanisms and saddle-points for these
conversions, namely $\textnormal{B}\rightarrow\textnormal{C}$ and
$\textnormal{D}\rightarrow\textnormal{E}$, are indicated between
the respective reaction steps. Again, the moving atoms are enclosed
within parentheses.

The minimum energy path for formation of $\mathrm{C_{i}O_{i}}$ is
clearly limited by the migration barrier of $\mathrm{C_{i}}$ and
not by a capture barrier. This is compatible with the observation
of $\mathrm{C_{i}O_{i}}$ in irradiated material at room-temperature.
Accordingly, the saddle point energy for

\begin{equation}
\mathrm{C_{i}}\textnormal{-}\mathrm{O_{i}}+\mathrm{O_{i}}\;(\textnormal{B})\stackrel{\mathrm{(C_{i})O_{i}+O_{i}}}{\longrightarrow}\mathrm{C_{i}}\mathrm{O_{i}}+\mathrm{O_{i}}\;(\textnormal{C}),\label{cr:bc1}
\end{equation}
is located at 0.18~eV in the energy scale, below the 0.76~eV of
the state involving the migration of $\mathrm{C_{i}}$. Obviously,
the large migration barrier of $\mathrm{O_{i}}$ inhibits the formation
of $\mathrm{C_{i}O_{i}}$ via migration of oxygen.

Note that the left reactant in Reaction~\ref{cr:bc1} is distinct
from the metastable $\mathrm{C}_{\mathrm{i}}\mathrm{O}_{\mathrm{i}}^{*}$
precursor reported in Ref.~\citep{Khirunenko2008}. In the $\mathrm{C_{i}}\textnormal{-}\mathrm{O_{i}}$
structure, both C and O atoms are separated by a Si-Si bond, but unlike
$\mathrm{C}_{\mathrm{i}}\mathrm{O}_{\mathrm{i}}^{*}$, they do not
share the same crystallographic plane. According to the energy scale
of Figure~\ref{fig2}, $\mathrm{C}_{\mathrm{i}}\mathrm{O}_{\mathrm{i}}^{*}$
is located at $-0.56$~eV. Direct conversion of $\mathrm{C}_{\mathrm{i}}\mathrm{O}_{\mathrm{i}}^{*}$
into $\mathrm{C_{i}}\mathrm{O_{i}}$ via in-plane jump of carbon or
oxygen atoms involves surmounting a barrier of at least 2.0~eV. Alternatively,
the transformation of $\mathrm{C}_{\mathrm{i}}\mathrm{O}_{\mathrm{i}}^{*}$
firstly into triclinic $\mathrm{C_{i}}\textnormal{-}\mathrm{O_{i}}$
(step B) followed by conversion into $\mathrm{C_{i}}\mathrm{O_{i}}$
(step C) has an overall barrier of only 0.94~eV. This is in excellent
agreement with the measurements of Abdullin \emph{et~al.} \citep{Abdullin1990},
who reported a 1~eV activation energy for the growth of $\mathrm{C_{i}}\mathrm{O_{i}}$
DLTS signal at the expense of another peak related to its precursor.

Dissociation of $\mathrm{C_{i}O_{i}}$ essentially consists of the
reversed formation mechanism (red line in Figure~\ref{fig2}). The
overall activation energy for dissociation, $E_{\textnormal{d}}=2.29$~eV,
is also governed by the migration of $\mathrm{C_{i}}$, which follows
from its off-plane detachment from oxygen in a first stage (reversal
of the reaction~\ref{cr:bc1}). This result slightly overestimates
the $\sim\!2$~eV from early measurements of the activation energy
for the annealing of $\mathrm{C_{i}O_{i}}$ \citep{Svensson1986a,Svensson1986}.

The alternative mechanism, where instead of $\mathrm{C_{i}}$ motion,
the first stage involves a detachment of $\mathrm{O_{i}}$ from $\mathrm{C_{i}O_{i}}$,

\begin{equation}
\mathrm{C_{i}}\mathrm{O_{i}}+\mathrm{O_{i}}\;(\textnormal{C})\stackrel{\mathrm{C_{i}(O_{i})+O_{i}}}{\longrightarrow}\mathrm{C_{i}}\textnormal{-}\mathrm{O_{i}}+\mathrm{O_{i}}\;(\textnormal{B}),\label{cr:dis}
\end{equation}
was also inspected. Reaction~\ref{cr:dis} starts from ground state
at step C (red) and has a barrier of 2.07~eV with saddle-point at
$+0.51$~eV shown in Figure~\ref{fig2} in black. However, since
$\mathrm{C_{i}}$ still has to escape from oxygen, the overall barrier
for dissociation is also $E_{\textnormal{d}}=2.29$~eV. This dissociation
route is nevertheless unlikely to occur due to a higher first stage
barrier.

Now we turn to $\mathrm{C_{i}O_{2i}}$. Its formation can occur either
(i) via further accumulation of oxygen in $\mathrm{C_{i}O_{i}}$,
which means starting with reactants $\mathrm{C_{i}O_{i}}+\mathrm{O_{i}}$
(step C, red line) in Figure~\ref{fig2}, or (ii) from reaction between
$\mathrm{C_{i}}$ and $\mathrm{O_{2i}}$, where the starting conditions
are represented by $\mathrm{C_{i}}+\mathrm{O_{2i}}$ (step C, blue
line). Of course, in Cz-Si, where the concentration of oxygen is much
larger than that of carbon, the initial state of option (ii) can only
be achieved upon release of $\mathrm{C_{i}}$ defects from the overwhelming
concentration of oxygen traps. Hence, option (ii) actually involves
two simultaneous reactions, namely (ii.1) the dissociation of $\mathrm{C_{i}O_{i}}$
(reaction $\textnormal{C}\rightarrow\textnormal{A}$) and (ii.2) capture
of $\mathrm{C_{i}}$ by $\mathrm{O_{2i}}$ to the more stable $\mathrm{C_{i}O_{2i}}$
complex (reaction $\textnormal{C}\rightarrow\textnormal{E}$). The
alternative formation mechanism implying the capture of a diffusing
$\mathrm{O_{2i}}$ by $\mathrm{C_{i}}$ is not physically probable
due to the high barrier involved.

According to Figure~\ref{fig2}, the formation mechanism (i) can
occur via migration of $\mathrm{O_{i}}$ with an overall barrier of
2.53~eV, or via migration of $\mathrm{C_{i}O_{i}}$ with an activation
energy of 2.45~eV. However, these reaction routes are shortcut by
dissociation of $\mathrm{C_{i}O_{i}}$ (which has an energy barrier
of 2.29~eV only), thus providing the necessary conditions for activation
of the formation mechanism (ii). Like $\mathrm{C_{i}O_{i}}$, the
formation of $\mathrm{C_{i}O_{2i}}$ via mechanism (ii) along the
blue line of Figure~\ref{fig2} is only limited by the migration
barrier of $\mathrm{C_{i}}$. However, because the first stage reaction
(ii.1) actually involves the dissociation of $\mathrm{C_{i}O_{i}}$
(reaction $\textnormal{C}\rightarrow\textnormal{A}$), the formation
mechanism of $\mathrm{C_{i}O_{2i}}$ is effectively activated by the
dissociation barrier of $\mathrm{C_{i}O_{i}}$, \emph{i.e.}, $E_{\textnormal{a}}=2.29$~eV.
Note that the state $(\mathrm{C_{i}})+\mathrm{O_{i}}+\mathrm{O_{i}}$
at 0.73~eV which limits the dissociation of $\mathrm{C_{i}O_{i}}$
to make carbon interstitials available, is higher in energy than $(\mathrm{C_{i}})+\mathrm{O_{2i}}$
at 0.40~eV governing the capture of $\mathrm{C_{i}}$ by $\mathrm{O_{2i}}$.
This picture explains the observed correlation between the dissociation
of $\mathrm{C_{i}O_{i}}$ and formation of $\mathrm{C_{i}O_{2i}}$
\citep{Ganagona2012,Raeissi2013,Ayedh2019}.

Regarding the dissociation of $\mathrm{C_{i}O_{2i}}$, we calculated
four possible scenarios which differ in the initial $\textnormal{E}\rightarrow\textnormal{D}$
step as depicted in Figure~\ref{fig2}, and can be summarized as
follows: (i) jump of $\mathrm{C_{i}}$ away from $\mathrm{C_{i}O_{2i}}$,
leaving $\mathrm{O_{2i}}$ behind; (2) detachment of $\mathrm{O_{i}}$
from $\mathrm{\mathrm{C_{i}O_{2i}}}$ leaving a $\mathrm{\mathrm{C_{i}O_{i}}}$
complex; (3) jump of $\mathrm{O_{2i}}$ away from $\mathrm{\mathrm{C_{i}O_{2i}}}$,
leaving a $\mathrm{\mathrm{C_{i}}}$ defect, and (4) detachment of
a $\mathrm{C_{i}O_{i}}$ complex from $\mathrm{\mathrm{C_{i}O_{2i}}}$,
thus leaving $\mathrm{O_{i}}$. Figure~\ref{fig2} clearly indicates
that mechanism (i) is the most favorable dissociation route, showing
an activation energy of $E_{\textnormal{d}}=3.06$~eV. This figure
is about 0.5~eV higher than what was recently observed by some of
us during annealing experiments \citep{Ayedh2019}. We can only reconcile
the calculations with the measurements if we assume that, instead
of a dissociation, the relevant annealing mechanism involves the capture
of interstitial oxygen by $\mathrm{C_{i}O_{2i}}$, namely

\begin{equation}
\mathrm{C_{i}}\mathrm{O_{2i}}+\mathrm{O_{i}}\stackrel{\mathrm{C_{i}}\mathrm{O_{2i}}+(\mathrm{O_{i}})}{\longrightarrow}\mathrm{C_{i}}\mathrm{O_{3i}},\label{cr:anneal-co3}
\end{equation}
thus explaining the measured activation energy $E_{\textnormal{a}}=2.55$~eV
for the annealing of $\mathrm{\mathrm{C_{i}O_{2i}}}$ \citep{Ayedh2019},
which is virtually identical to the migration barrier of $\mathrm{O_{i}}$.
We found that the formation of $\mathrm{C_{i}}\mathrm{O_{3i}}$ according
to Reaction~\ref{cr:anneal-co3} is energetically favorable and corresponds
to a lowering of the energy by $E_{\textrm{b}}=0.76$~eV. Here the
ground state structure of $\mathrm{C_{i}O_{3i}}$ consisted of a staggered
oxygen trimer next to $\mathrm{C_{i}}$. The binding energy of $\mathrm{C_{i}}$
to $\mathrm{O_{3i}}$ was calculated as $E_{\textnormal{b}}=2.56$~eV.
This figure follows the trend shown by $E_{\textnormal{b}}=2.33$~eV
and 1.56~eV as obtained for the analogous quantity regarding the
attachment of $\mathrm{C_{i}}$ to $\mathrm{O_{2i}}$ and $\mathrm{O_{i}}$,
respectively (see Reactions~\ref{cr:eb-cioi} and \ref{cr:eb-cio2i}).
The proposed mechanism as described by Reaction~\ref{cr:anneal-co3}
is also compliant with the observed first order kinetics of the annealing
--- the huge concentration of $\mathrm{O_{i}}$ is effectively invariant
during the process.

In Ref.~\citep{Ayedh2019}, an attempt rate for the annealing kinetics
of $\mathrm{\mathrm{C_{i}O_{2i}}}$ was measured as $\nu_{\infty}=(2\textnormal{-}30)\times10^{12}\,\textnormal{s}^{-1}$.
This figure is in line with the Debye frequency of Si, suggesting
that the annealing should be prompted by atomic vibrations. However,
our proposal, according to which $\mathrm{\mathrm{C_{i}O_{2i}}}$
anneals out due to capture of mobile $\mathrm{O_{i}}$ impurities
is challenged by the fact that $\nu_{\infty}$ is two orders of magnitude
slower than the analogous figure obtained for $\mathrm{O_{i}}$ diffusivity
in silicon \citep{Newman2000}. The latter is obviously too high to
be described as a simple phonon-assisted jump and, as far as we know,
there is no clear explanation for such anomaly. At the moment we can
only infer that the forward rate of Reaction 8 is limited by a phonon-assisted
process, and therefore governed by a physics that somehow differs
from that of the migration of isolated $\mathrm{O_{i}}$. Hence, while
the activation energy for the annealing of $\mathrm{\mathrm{C_{i}O_{2i}}}$
corresponds to the large migration barrier of oxygen at remote locations
from $\mathrm{\mathrm{C_{i}O_{2i}}}$, the measured attempt frequency
in the $10^{13}$-Hz range may simply reflect the kinetics of the
slower final steps, when both impurities are in close proximity, which
cannot be described as isolated $\mathrm{O_{i}}$ jumps anymore, but
rather as a restructuring of a $\mathrm{\mathrm{C_{i}O_{3i}}}$ complex.

We end this section with a few considerations on the charge state
dependence of formation and dissociation of $\mathrm{\mathrm{C_{i}O_{i}}}$
and $\mathrm{\mathrm{C_{i}O_{2i}}}$ complexes. Firstly, the above
results refer to neutral defects and essentially they are valid for
intrinsic material or doped-Si subject to annealing ($T\sim300\textnormal{-}500\,{}^{\circ}$C).

Secondly, we note that the formation and dissociation mechanisms do
not involve long range Coulomb interactions between reactants. Hence,
in p-type or n-type Si, the energies of the stable states in steps
A, C and E should be lowered by an amount that corresponds to the
depth of the hole or electron traps of $\mathrm{C_{i}}$, $\mathrm{C_{i}O_{i}}$,
or $\mathrm{C_{i}O_{2i}}$. Analogously, saddle-point energies should
consider a charge state effect. In this case, both `red' and `blue'
MEPs of Figure~\ref{fig2} are limited by the migration barrier of
$\mathrm{C_{i}}$. Therefore, if the thermodynamic conditions are
such that the Fermi level is above $E_{\textnormal{c}}-0.12$~eV
\citep{Song1990} or below $E_{\textnormal{v}}+0.28$~eV \citep{Lee1977},
the $\mathrm{C_{i}}$ defect is negatively or positively charged,
so that migration barriers for $\mathrm{C_{i}^{-}}$ or $\mathrm{C_{i}^{+}}$
should be considered, respectively. For instance, in heavily-doped
p-type Si, the migration barrier of $\mathrm{C_{i}^{+}}$ was measured
as $E_{\mathrm{m}}=0.89$~eV. Considering the hole trap energies
of $\mathrm{C_{i}}$ ($E_{\textnormal{v}}+0.28$~eV) and $\mathrm{C_{i}O_{i}}$
($E_{\textnormal{v}}+0.36$~eV), we estimate activation energies
for formation and dissociation of $\mathrm{C_{i}O_{i}}$ of $E_{\textnormal{a}}=0.89$~eV
and $E_{\textnormal{a}}=2.53$~eV, respectively. These figures are
slightly larger than the analogous quantities reported above for neutral
defects (0.73~eV and 2.29~eV, respectively).

\subsection{Electronic properties}

\begin{figure}
\includegraphics[width=7cm]{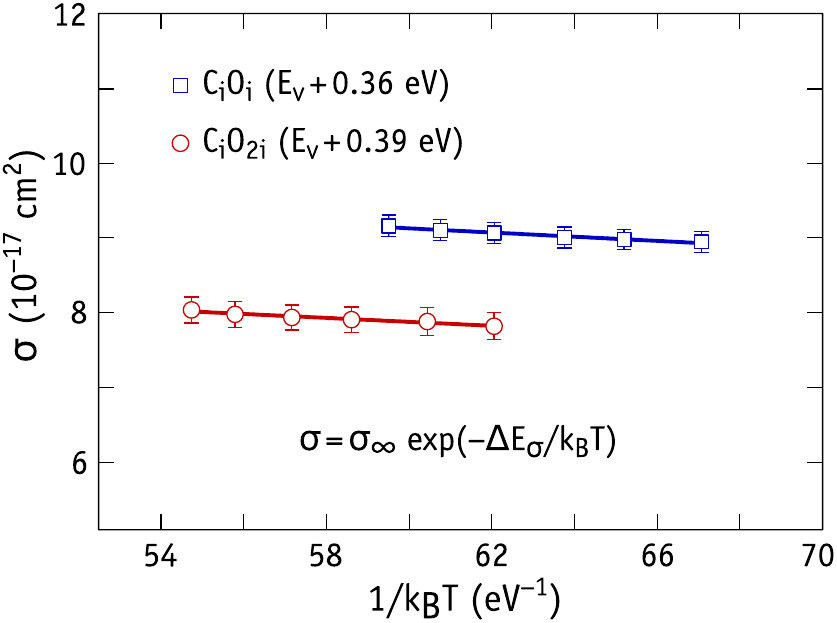}

\caption{\label{fig3} Temperature dependence of the hole capture cross section
($\sigma$) of $\mathrm{C_{i}O_{i}}$ and $\mathrm{C_{i}O_{2i}}$
traps. $\sigma(T)$ for hole capture by neutral $\mathrm{C_{i}O_{i}}$
and $\mathrm{C_{i}O_{2i}}$ shows very week temperature dependency
due to the very small capture barrier for both complexes (as extracted
from the slopes of linear fittings). $E_{\textrm{v}}$, $\sigma_{\infty}$,
and $\Delta E_{\sigma}$ stand for the valence band top energy, the
direct hole capture cross section and respective thermally activated
capture barrier. Straight lines are linear fits of $\sigma(T)$ to
the data.}
\end{figure}

In a recent study, \{C,O\}-rich Si samples were irradiation at room-temperature
and annealed at $400~^{\circ}$C for 30~h in order to anneal out
$\mathrm{C_{i}O_{i}}$ and form the $\mathrm{C_{i}O_{2i}}$ \citep{Ayedh2019}.
The evolution of the defects was monitored by DLTS via observation
of the corresponding hole traps with activation energy for hole emission
of $\Delta E_{\textnormal{h}}=0.36$~eV and $\Delta E_{\textnormal{h}}=0.39$~eV.
In order to obtain the thermodynamic transition levels, a capture
barrier has to be subtracted from $\Delta E_{\textnormal{h}}$ values,
and after that we can compare the observed transitions with corresponding
calculations based on ground state energies. Below we present results
from measurements of the capture barriers of $\mathrm{C_{i}O_{i}}$
and $\mathrm{C_{i}O_{2i}}$, which supplement the study of Ref.~\citep{Ayedh2019}.

The hole capture cross section ($\sigma$) of $\mathrm{C_{i}O_{i}}$
and $\mathrm{C_{i}O_{2i}}$ was measured at different temperatures
and compared for both complexes. Values of $\mathrm{\sigma}$ as a
function of the inverse of the absolute temperature are plotted in
Fig.~\ref{fig3} . The $\mathrm{\sigma}$ values were extracted from
the observed amplitude of the DLTS peak as a function of the filling
pulse duration (between 10~ns and 10~μs) at sample temperatures
in the range 170-215~K. For very short pulses (10-300~ns), the number
of filled $\mathrm{C_{i}O_{i}}$ and $\mathrm{C_{i}O_{2i}}$ traps
was negligible. On the other hand, for pulses longer than 3~μs both
signals saturated due to complete filling of the traps. The two levels
exhibit close capture cross sections ($\thicksim\!9\times10^{-17}\:\mathrm{cm^{2}}$
for $\mathrm{C_{i}O_{i}}$ versus $\sim\!8\times10^{-17}\:\mathrm{cm^{2}}$
for $\mathrm{C_{i}O_{2i}}$) with very weak impact of the temperature
variation. However, $\sigma$ is a temperature-dependent quantity
which can be described as,

\begin{equation}
\sigma=\sigma_{\infty}\,\exp(-\Delta E_{\sigma}/k_{\textnormal{B}}T),\label{cr:notation3}
\end{equation}
where $\sigma_{\infty}$ is the direct capture cross section (high
temperature limit), $\Delta E_{\sigma}$ is the thermally activated
hole capture barrier and $k_{\textnormal{B}}$ the Boltzmann constant.
Values of $\Delta E_{\sigma}$ and $\sigma_{\infty}$ for $\mathrm{C_{i}O_{i}}$
and $\mathrm{C_{i}O_{2i}}$ were extracted by fitting Eq.~\ref{cr:notation3}
to the measured data as depicted in Fig.~\ref{fig3}. We found that
$\mathrm{\Delta E_{\sigma}}$ is very small and almost identical for
both defects, $\sim\!3$~meV for $\mathrm{C_{i}O_{i}}$ and $\sim\!4$~meV
for $\mathrm{C_{i}O_{2i}}$. This result is in line with the model
where the $\mathrm{C_{i}O_{2i}}$ hole trap can be described as a
$\mathrm{C_{i}O_{i}}$ trap perturbed by the presence of a nearby
interstitial oxygen atom.

Activation energies of hole emission and apparent capture cross sections
for $\mathrm{C_{i}O_{i}}$ and $\mathrm{C_{i}O_{2i}}$ as reported
in Ref.~\citep{Ayedh2019} are listed in Table \ref{tab1}. These
data are also accompanied by direct capture cross sections and capture
barriers for both complexes as obtained from the present measurements.
In addition, the measured and calculated (see below) electronic levels
$[E(0/+)-E_{\textnormal{v}}]$ of $\mathrm{C_{i}O_{i}}$ and $\mathrm{C_{i}O_{2i}}$
are also included.

\begin{table}
\caption{\label{tab1} Activation energies for hole emission ($\Delta E_{\textnormal{h}}$)
and apparent capture cross sections ($\sigma_{\textnormal{a}}$) (from
Ref.~\citep{Ayedh2019}), direct capture cross sections ($\sigma_{\infty}$)
and capture barriers ($\Delta E_{\sigma}$) (this work) of $\mathrm{C_{i}O_{i}}$
and $\mathrm{C_{i}O_{2i}}$ complexes in Si. Capture cross sections
and energies are given in cm$^{2}$ and eV, respectively. Measured
and calculated donor levels, $E(0/+)-E_{\textrm{v}}$, are also reported.}

\begin{ruledtabular}
\centering{}%
\begin{tabular}{lcccccc}
 & \multicolumn{4}{c}{Hole trap data} & \multicolumn{2}{c}{$E(0/+)-E_{\textnormal{v}}$}\tabularnewline
\hline 
 & $\Delta E_{\textnormal{h}}$ & $\sigma_{\textnormal{a}}$ & $\sigma_{\infty}$ & $\Delta E_{\sigma}$ & Measured & Calculated\tabularnewline
$\mathrm{\mathrm{C_{i}O_{i}}}$ & 0.36 & $1\times10^{-15}$ & $1.1\times10^{-16}$ & 0.003 & 0.36 & 0.30\tabularnewline
$\mathrm{\mathrm{C_{i}O_{2i}}}$ & 0.39 & $2\times10^{-15}$ & $9.6\times10^{-17}$ & 0.004 & 0.39 & 0.33\tabularnewline
\end{tabular}
\end{ruledtabular}

\end{table}

The capture barriers for $\mathrm{C_{i}O_{i}}$ and $\mathrm{C_{i}O_{2i}}$
are very small, and therefore, the measured activation energies for
hole emission essentially represent the location of the donor transition
with respect to the valence band top. Furthermore, no variation was
found for the hole emission rate of both $\mathrm{C_{i}O_{i}}$ and
$\mathrm{C_{i}O_{2i}}$ traps with the application of different electric
fields during the DLTS experiments. The lack of a Poole-Frenkel effect
indicates that both levels are likely to correspond to donor transitions.

A transition level between two charge states, say $q$ and $q'$,
of a defect (with $q$ being more negative than $q'$) is defined
as

\begin{eqnarray}
E(q/q') & = & -\frac{E^{(q)}(R)-E^{(q')}(R')}{q-q'},\label{eq:level}
\end{eqnarray}
where $E^{(q)}$ is the energy of the supercell with the defect in
charge state $q$ and $R$ a generalized coordinate representative
of the defect configuration. Equation~(\ref{eq:level}) accounts
for the fact that charge states $q$ and $q'$ may correspond to radically
different atomistic geometries $R$ and $R'$, respectively (which
is not the present case). The use of periodic boundary conditions
imply that the supercell is always neutral irrespectively of the number
of electrons in the system. To mitigate this spurious effect, energies
in Eq.~(\ref{eq:level}) are off-set by a periodic charge correction
according to Freysoldt \emph{et~al.} \citep{Freysoldt2009}.

In order to cast the levels in a way that they can be compared to
the experiments, \emph{i.e.}, $E(0/+)-E_{\textnormal{v}}$, we have
to calculate the energy of the valence band top. This is done by using
Eq.~(\ref{eq:level}) for the case of a bulk (defect-free) supercell,
$E_{\textnormal{v}}=E_{\textnormal{bulk}}(0/+)$.

The calculated donor levels of $\mathrm{C_{i}O_{i}}$ and $\mathrm{C_{i}O_{2i}}$
are shown in the right-most column of Table~\ref{tab1}. They are
underestimated with respect to the observations by 60~meV, but significantly,
they account for the observed relative depth, i.e., the donor transition
of $\mathrm{C_{i}O_{2i}}$ is 30~meV above the donor transition of
$\mathrm{C_{i}O_{i}}$.
\noindent \begin{center}
\begin{figure}
\begin{centering}
\includegraphics[width=7cm]{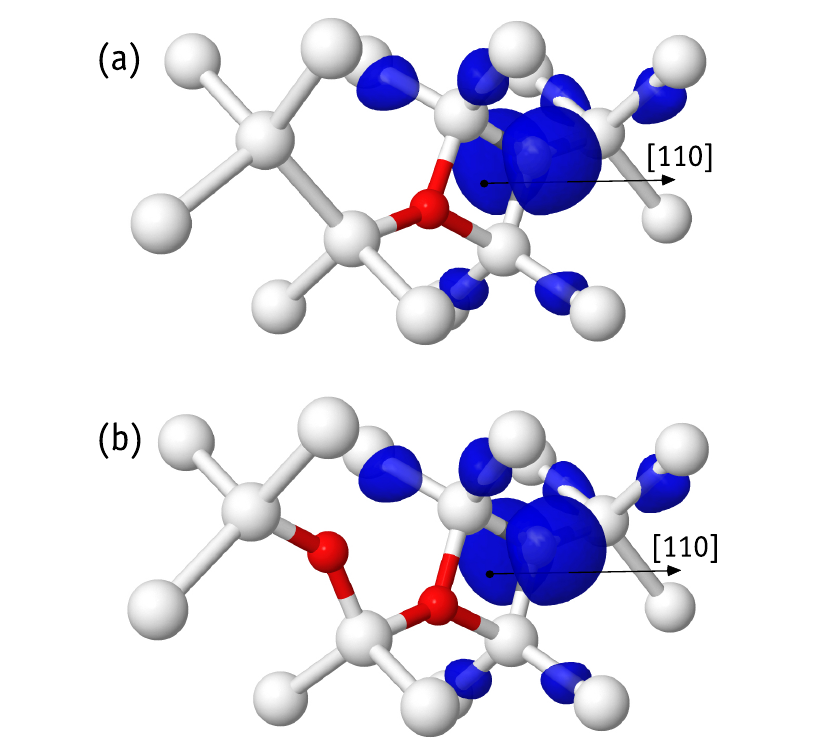}
\par\end{centering}
\caption{\label{fig:density}Isosurface of the electron density related to
the highest occupied state (donor state) of (a) $\mathrm{C_{i}O_{i}}$
and (b) $\mathrm{C_{i}O_{2i}}$. The isosurface cut-off value is identical
for both cases, $\rho_{\textrm{cut}}=0.001$~Bohr$^{-3}$. The view-point
is analogous to that of Fig.~\ref{fig1}(c). The $[110]$ crystallographic
direction is perpendicular to the mirror plane of the defects.}
\end{figure}
\par\end{center}

The hole traps arise from a fully occupied p-like state centered on
the carbon atom, and lying deep in the band gap. Figure~\ref{fig:density}
depicts the donor state (Kohn-Sham state) within the gap in the form
of an electron density isosurface (blue), related to the highest occupied
state of neutral $\mathrm{C_{i}O_{i}}$ and $\mathrm{C_{i}O_{2i}}$
defects. The carbon atom is hidden by the large $p$-orbital-like
shape near the arrow that indicates the $[110]$ direction. There
is no apparent difference between the two defects. This is consistent
with the nearly identical characteristics determined for both, namely
the activation energy for hole emission, the capture barriers and
capture cross sections.

The fact that the donor level of $\mathrm{C_{i}O_{2i}}$ is slightly
higher (30~meV) in the gap, can be explained by the repulsion of
the donor electrons on the $p$-state by the additional and highly
electronegative O atom. This effect is analogous to that invoked to
explain the raise of the donor transitions of the thermal double donors
in Si and Ge, where the increasing number of oxygen atoms accumulated
on each donor leads to progressively shallower levels \citep{Coutinho2001a}.

\section{Conclusions\label{sec:conclusions}}

We presented a detailed model regarding the formation and dissociation
mechanisms of carbon-oxygen complexes, involving interstitial carbon
and interstitial oxygen impurities. The results are based on hybrid
density functional theory. We also supplement previous experimental
data with capture kinetics measurements, which allowed us to estimate
and compare the capture barriers of $\mathrm{C_{i}O_{i}}$ and $\mathrm{C_{i}O_{2i}}$.

Like $\mathrm{C_{i}O_{i}}$, the $\mathrm{C_{i}O_{2i}}$ complex is
made of a square-like structure involving C, O and two Si atoms. Both
defects show large binding energies,

\begin{equation}
\mathrm{C_{i}}+\mathrm{O_{i}}+\mathrm{O_{i}}\stackrel{E_{\textnormal{m}}=0.73\,\textnormal{eV}}{\longrightarrow}\mathrm{C_{i}}\mathrm{O_{i}}+\mathrm{O_{i}}+[E_{\textnormal{b}}=1.56\,\textnormal{eV}],\label{eq:conclusions1}
\end{equation}

\begin{equation}
\mathrm{C_{i}}+\mathrm{O_{2i}}\stackrel{E_{\textnormal{m}}=0.73\,\textnormal{eV}}{\longrightarrow}\mathrm{C_{i}O_{2i}}+[E_{\textnormal{b}}=2.33\,\textnormal{eV}],\label{eq:conclusions2}
\end{equation}
which are suggestive of a high thermal stability. The kinetics of
both reactions above are thermally activated by the migration barrier
of $\mathrm{C_{i}}$, ($E_{\textnormal{m}}=0.73$~eV). However, in
O-rich material and below the annealing temperature of $\mathrm{C_{i}O_{i}}$
($\sim\!400~^{\circ}$C), Reaction~\ref{eq:conclusions1} will dominate
due to the large concentration $\mathrm{O_{i}}$ traps for the fast-diffusing
$\mathrm{C_{i}}$ (in comparison to the concentration of $\mathrm{O_{2i}}$).

Dissociation of $\mathrm{C_{i}O_{i}}$ is simply governed by Reaction~\ref{eq:conclusions1}
in the backwards direction. Now the carbon detaches and escapes from
$\mathrm{O_{i}}$. The calculated overall barrier is $E_{\textnormal{d}}=1.56+0.73=2.29$~eV.
This result implies that $\mathrm{C_{i}O_{i}}$ should dissociate
at a slightly lower temperature than that for migration of $\mathrm{O_{i}}$
(with an activation barrier of about 2.5~eV).

As referred above, around room temperature, $\mathrm{C_{i}}$ defects
become mobile and are quickly consumed by abundant $\mathrm{O_{i}}$
to form $\mathrm{C_{i}O_{i}}$. The $\mathrm{C_{i}O_{2i}}$ is observed
only after (1) annealing out $\mathrm{C_{i}O_{i}}$ by reversing Reaction~\ref{eq:conclusions1},
and (2) the released $\mathrm{C_{i}}$ impurities travel and find
$\mathrm{O_{2i}}$, according to Reaction~\ref{eq:conclusions2},
ending up in a more stable complex ($E_{\mathrm{b}}=2.33$~eV). This
process is limited by step 1 (annealing of $\mathrm{C_{i}O_{i}}$),
meaning that it thermally activated by the dissociation barrier of
$\mathrm{C_{i}O_{i}}$ ($E_{\textnormal{d}}=2.29$~eV). Again, this
happens at temperatures below the threshold for migration of $\mathrm{O_{i}}$.

The annealing of $\mathrm{C_{i}O_{2i}}$ was found to follow from
the capture of mobile $\mathrm{O_{i}}$ impurities by $\mathrm{C_{i}O_{2i}}$
as,

\begin{equation}
\mathrm{C_{i}}\mathrm{O_{2i}}+\mathrm{O_{i}}\stackrel{E_{\textnormal{m}}=2.53\,\textnormal{eV}}{\longrightarrow}\mathrm{C_{i}}\mathrm{O_{3i}},
\end{equation}
whose kinetics is limited by the migration barrier of interstitial
oxygen impurities ($E_{\textnormal{m}}=2.53\,\textnormal{eV}$). This
picture accounts well for the annealing measurements of $\mathrm{C_{i}O_{2i}}$,
which was found to be thermally activated by a barrier $E_{\textnormal{a}}=2.55$~eV.
Dissociation of $\mathrm{C_{i}O_{2i}}$ into smaller moieties had
barriers invariably above 3~eV.

The electronic properties of $\mathrm{C_{i}O_{i}}$ and $\mathrm{C_{i}O_{2i}}$
were compared by calculating their donor levels, as well as measuring
their respective capture cross sections. Both theory and experiments
converge well --- $\mathrm{C_{i}O_{i}}$ has a donor transition at
$E_{\textnormal{c}}+0.36$~eV, only 30~meV below the analogous transition
of $\mathrm{C_{i}O_{2i}}$.
\begin{acknowledgments}
Financial support by the Norwegian Research Council through the research
project OxSil (no. 254977) is gratefully acknowledged. The Research
Council of Norway is also acknowledged for the support to the Norwegian
Micro- and Nano-Fabrication Facility, NorFab, (project no. 245963).
HMA thanks the Faculty of Mathematics and Natural Sciences / University
of Oslo - Norway for the Kristine Bonnevie travel grant to support
his research visit to University of Aveiro. JC thanks the support
of the i3N project, Refs. UIDB/50025/2020 and UIDP/50025/2020, financed
by the Fundação para a Ciência e a Tecnologia in Portugal.
\end{acknowledgments}

\bibliographystyle{apsrev4-1}

\begin{thebibliography}{72}%
\makeatletter
\providecommand \@ifxundefined [1]{%
 \@ifx{#1\undefined}
}%
\providecommand \@ifnum [1]{%
 \ifnum #1\expandafter \@firstoftwo
 \else \expandafter \@secondoftwo
 \fi
}%
\providecommand \@ifx [1]{%
 \ifx #1\expandafter \@firstoftwo
 \else \expandafter \@secondoftwo
 \fi
}%
\providecommand \natexlab [1]{#1}%
\providecommand \enquote  [1]{``#1''}%
\providecommand \bibnamefont  [1]{#1}%
\providecommand \bibfnamefont [1]{#1}%
\providecommand \citenamefont [1]{#1}%
\providecommand \href@noop [0]{\@secondoftwo}%
\providecommand \href [0]{\begingroup \@sanitize@url \@href}%
\providecommand \@href[1]{\@@startlink{#1}\@@href}%
\providecommand \@@href[1]{\endgroup#1\@@endlink}%
\providecommand \@sanitize@url [0]{\catcode `\\12\catcode `\$12\catcode
  `\&12\catcode `\#12\catcode `\^12\catcode `\_12\catcode `\%12\relax}%
\providecommand \@@startlink[1]{}%
\providecommand \@@endlink[0]{}%
\providecommand \url  [0]{\begingroup\@sanitize@url \@url }%
\providecommand \@url [1]{\endgroup\@href {#1}{\urlprefix }}%
\providecommand \urlprefix  [0]{URL }%
\providecommand \Eprint [0]{\href }%
\providecommand \doibase [0]{http://dx.doi.org/}%
\providecommand \selectlanguage [0]{\@gobble}%
\providecommand \bibinfo  [0]{\@secondoftwo}%
\providecommand \bibfield  [0]{\@secondoftwo}%
\providecommand \translation [1]{[#1]}%
\providecommand \BibitemOpen [0]{}%
\providecommand \bibitemStop [0]{}%
\providecommand \bibitemNoStop [0]{.\EOS\space}%
\providecommand \EOS [0]{\spacefactor3000\relax}%
\providecommand \BibitemShut  [1]{\csname bibitem#1\endcsname}%
\let\auto@bib@innerbib\@empty
\bibitem [{\citenamefont {Kaiser}\ \emph {et~al.}(1958)\citenamefont {Kaiser},
  \citenamefont {Frisch},\ and\ \citenamefont {Reiss}}]{Kaiser1958}%
  \BibitemOpen
  \bibfield  {author} {\bibinfo {author} {\bibfnamefont {W.}~\bibnamefont
  {Kaiser}}, \bibinfo {author} {\bibfnamefont {H.~L.}\ \bibnamefont {Frisch}},
  \ and\ \bibinfo {author} {\bibfnamefont {H.}~\bibnamefont {Reiss}},\ }\href
  {\doibase 10.1103/physrev.112.1546} {\bibfield  {journal} {\bibinfo
  {journal} {Physical Review}\ }\textbf {\bibinfo {volume} {112}},\ \bibinfo
  {pages} {1546} (\bibinfo {year} {1958})}\BibitemShut {NoStop}%
\bibitem [{\citenamefont {Bean}\ and\ \citenamefont {Newman}(1972)}]{Bean1972}%
  \BibitemOpen
  \bibfield  {author} {\bibinfo {author} {\bibfnamefont {A.~R.}\ \bibnamefont
  {Bean}}\ and\ \bibinfo {author} {\bibfnamefont {R.~C.}\ \bibnamefont
  {Newman}},\ }\href {\doibase 10.1016/0022-3697(72)90004-2} {\bibfield
  {journal} {\bibinfo  {journal} {Journal of Physics and Chemistry of Solids}\
  }\textbf {\bibinfo {volume} {33}},\ \bibinfo {pages} {255} (\bibinfo {year}
  {1972})}\BibitemShut {NoStop}%
\bibitem [{\citenamefont {Schmidt}(2003)}]{Schmidt2003}%
  \BibitemOpen
  \bibfield  {author} {\bibinfo {author} {\bibfnamefont {J.}~\bibnamefont
  {Schmidt}},\ }\href {\doibase 10.4028/www.scientific.net/ssp.95-96.187}
  {\bibfield  {journal} {\bibinfo  {journal} {Solid State Phenomena}\ }\textbf
  {\bibinfo {volume} {95-96}},\ \bibinfo {pages} {187} (\bibinfo {year}
  {2003})}\BibitemShut {NoStop}%
\bibitem [{\citenamefont {Schmidt}\ and\ \citenamefont
  {Bothe}(2004)}]{Schmidt2004}%
  \BibitemOpen
  \bibfield  {author} {\bibinfo {author} {\bibfnamefont {J.}~\bibnamefont
  {Schmidt}}\ and\ \bibinfo {author} {\bibfnamefont {K.}~\bibnamefont
  {Bothe}},\ }\href {\doibase 10.1103/physrevb.69.024107} {\bibfield  {journal}
  {\bibinfo  {journal} {Physical Review B}\ }\textbf {\bibinfo {volume} {69}},\
  \bibinfo {pages} {024107} (\bibinfo {year} {2004})}\BibitemShut {NoStop}%
\bibitem [{\citenamefont {Vaqueiro-Contreras}\ \emph
  {et~al.}(2019)\citenamefont {Vaqueiro-Contreras}, \citenamefont {Markevich},
  \citenamefont {Coutinho}, \citenamefont {Santos}, \citenamefont {Crowe},
  \citenamefont {Halsall}, \citenamefont {Hawkins}, \citenamefont {Lastovskii},
  \citenamefont {Murin},\ and\ \citenamefont
  {Peaker}}]{Vaqueiro-Contreras2019}%
  \BibitemOpen
  \bibfield  {author} {\bibinfo {author} {\bibfnamefont {M.}~\bibnamefont
  {Vaqueiro-Contreras}}, \bibinfo {author} {\bibfnamefont {V.~P.}\ \bibnamefont
  {Markevich}}, \bibinfo {author} {\bibfnamefont {J.}~\bibnamefont {Coutinho}},
  \bibinfo {author} {\bibfnamefont {P.}~\bibnamefont {Santos}}, \bibinfo
  {author} {\bibfnamefont {I.~F.}\ \bibnamefont {Crowe}}, \bibinfo {author}
  {\bibfnamefont {M.~P.}\ \bibnamefont {Halsall}}, \bibinfo {author}
  {\bibfnamefont {I.}~\bibnamefont {Hawkins}}, \bibinfo {author} {\bibfnamefont
  {S.~B.}\ \bibnamefont {Lastovskii}}, \bibinfo {author} {\bibfnamefont
  {L.~I.}\ \bibnamefont {Murin}}, \ and\ \bibinfo {author} {\bibfnamefont
  {A.~R.}\ \bibnamefont {Peaker}},\ }\href {\doibase 10.1063/1.5091759}
  {\bibfield  {journal} {\bibinfo  {journal} {Journal of Applied Physics}\
  }\textbf {\bibinfo {volume} {125}},\ \bibinfo {pages} {185704} (\bibinfo
  {year} {2019})}\BibitemShut {NoStop}%
\bibitem [{\citenamefont {Christopoulos}\ \emph {et~al.}(2016)\citenamefont
  {Christopoulos}, \citenamefont {Parfitt}, \citenamefont {Sgourou},
  \citenamefont {Londos}, \citenamefont {Vovk},\ and\ \citenamefont
  {Chroneos}}]{Christopoulos2016}%
  \BibitemOpen
  \bibfield  {author} {\bibinfo {author} {\bibfnamefont {S.-R.~G.}\
  \bibnamefont {Christopoulos}}, \bibinfo {author} {\bibfnamefont {D.~C.}\
  \bibnamefont {Parfitt}}, \bibinfo {author} {\bibfnamefont {E.~N.}\
  \bibnamefont {Sgourou}}, \bibinfo {author} {\bibfnamefont {C.~A.}\
  \bibnamefont {Londos}}, \bibinfo {author} {\bibfnamefont {R.~V.}\
  \bibnamefont {Vovk}}, \ and\ \bibinfo {author} {\bibfnamefont
  {A.}~\bibnamefont {Chroneos}},\ }\href {\doibase 10.1007/s10854-016-5249-z}
  {\bibfield  {journal} {\bibinfo  {journal} {Journal of Materials Science:
  Materials in Electronics}\ }\textbf {\bibinfo {volume} {27}},\ \bibinfo
  {pages} {11268} (\bibinfo {year} {2016})}\BibitemShut {NoStop}%
\bibitem [{\citenamefont {Newman}(2000)}]{Newman2000}%
  \BibitemOpen
  \bibfield  {author} {\bibinfo {author} {\bibfnamefont {R.~C.}\ \bibnamefont
  {Newman}},\ }\href {\doibase 10.1088/0953-8984/12/25/201} {\bibfield
  {journal} {\bibinfo  {journal} {Journal of Physics: Condensed Matter}\
  }\textbf {\bibinfo {volume} {12}},\ \bibinfo {pages} {R335} (\bibinfo {year}
  {2000})}\BibitemShut {NoStop}%
\bibitem [{\citenamefont {Watkins}\ and\ \citenamefont
  {Corbett}(1961)}]{Watkins1961}%
  \BibitemOpen
  \bibfield  {author} {\bibinfo {author} {\bibfnamefont {G.~D.}\ \bibnamefont
  {Watkins}}\ and\ \bibinfo {author} {\bibfnamefont {J.~W.}\ \bibnamefont
  {Corbett}},\ }\href {\doibase 10.1103/physrev.121.1001} {\bibfield  {journal}
  {\bibinfo  {journal} {Physical Review}\ }\textbf {\bibinfo {volume} {121}},\
  \bibinfo {pages} {1001} (\bibinfo {year} {1961})}\BibitemShut {NoStop}%
\bibitem [{\citenamefont {Lee}\ and\ \citenamefont {Corbett}(1976)}]{Lee1976}%
  \BibitemOpen
  \bibfield  {author} {\bibinfo {author} {\bibfnamefont {Y.-H.}\ \bibnamefont
  {Lee}}\ and\ \bibinfo {author} {\bibfnamefont {J.~W.}\ \bibnamefont
  {Corbett}},\ }\href {\doibase 10.1103/physrevb.13.2653} {\bibfield  {journal}
  {\bibinfo  {journal} {Physical Review B}\ }\textbf {\bibinfo {volume} {13}},\
  \bibinfo {pages} {2653} (\bibinfo {year} {1976})}\BibitemShut {NoStop}%
\bibitem [{\citenamefont {Watkins}(1965)}]{Watkins1965}%
  \BibitemOpen
  \bibfield  {author} {\bibinfo {author} {\bibfnamefont {G.~D.}\ \bibnamefont
  {Watkins}},\ }\enquote {\bibinfo {title} {Radiation damage in
  semiconductors},}\ \ (\bibinfo  {publisher} {Dunod, Paris},\ \bibinfo {year}
  {1965})\BibitemShut {NoStop}%
\bibitem [{\citenamefont {Watkins}\ and\ \citenamefont
  {Brower}(1976)}]{Watkins1976}%
  \BibitemOpen
  \bibfield  {author} {\bibinfo {author} {\bibfnamefont {G.~D.}\ \bibnamefont
  {Watkins}}\ and\ \bibinfo {author} {\bibfnamefont {K.~L.}\ \bibnamefont
  {Brower}},\ }\href {\doibase 10.1103/physrevlett.36.1329} {\bibfield
  {journal} {\bibinfo  {journal} {Physical Review Letters}\ }\textbf {\bibinfo
  {volume} {36}},\ \bibinfo {pages} {1329} (\bibinfo {year}
  {1976})}\BibitemShut {NoStop}%
\bibitem [{\citenamefont {Lee}\ \emph {et~al.}(1977)\citenamefont {Lee},
  \citenamefont {Cheng}, \citenamefont {Gerson}, \citenamefont {Mooney},\ and\
  \citenamefont {Corbett}}]{Lee1977}%
  \BibitemOpen
  \bibfield  {author} {\bibinfo {author} {\bibfnamefont {Y.~H.}\ \bibnamefont
  {Lee}}, \bibinfo {author} {\bibfnamefont {L.~J.}\ \bibnamefont {Cheng}},
  \bibinfo {author} {\bibfnamefont {J.~D.}\ \bibnamefont {Gerson}}, \bibinfo
  {author} {\bibfnamefont {P.~M.}\ \bibnamefont {Mooney}}, \ and\ \bibinfo
  {author} {\bibfnamefont {J.~W.}\ \bibnamefont {Corbett}},\ }\href {\doibase
  10.1016/0038-1098(77)91489-2} {\bibfield  {journal} {\bibinfo  {journal}
  {Solid State Communications}\ }\textbf {\bibinfo {volume} {21}},\ \bibinfo
  {pages} {109} (\bibinfo {year} {1977})}\BibitemShut {NoStop}%
\bibitem [{\citenamefont {Leary}\ \emph {et~al.}(1997)\citenamefont {Leary},
  \citenamefont {Jones}, \citenamefont {\"Oberg},\ and\ \citenamefont
  {Torres}}]{Leary1997}%
  \BibitemOpen
  \bibfield  {author} {\bibinfo {author} {\bibfnamefont {P.}~\bibnamefont
  {Leary}}, \bibinfo {author} {\bibfnamefont {R.}~\bibnamefont {Jones}},
  \bibinfo {author} {\bibfnamefont {S.}~\bibnamefont {\"Oberg}}, \ and\
  \bibinfo {author} {\bibfnamefont {V.~J.~B.}\ \bibnamefont {Torres}},\ }\href
  {\doibase 10.1103/physrevb.55.2188} {\bibfield  {journal} {\bibinfo
  {journal} {Physical Review B}\ }\textbf {\bibinfo {volume} {55}},\ \bibinfo
  {pages} {2188} (\bibinfo {year} {1997})}\BibitemShut {NoStop}%
\bibitem [{\citenamefont {Londos}(1987)}]{Londos1987}%
  \BibitemOpen
  \bibfield  {author} {\bibinfo {author} {\bibfnamefont {C.~A.}\ \bibnamefont
  {Londos}},\ }\href {\doibase 10.1103/physrevb.35.6295} {\bibfield  {journal}
  {\bibinfo  {journal} {Physical Review B}\ }\textbf {\bibinfo {volume} {35}},\
  \bibinfo {pages} {6295} (\bibinfo {year} {1987})}\BibitemShut {NoStop}%
\bibitem [{\citenamefont {Tipping}\ and\ \citenamefont
  {Newman}(1987)}]{Tipping1987}%
  \BibitemOpen
  \bibfield  {author} {\bibinfo {author} {\bibfnamefont {A.~K.}\ \bibnamefont
  {Tipping}}\ and\ \bibinfo {author} {\bibfnamefont {R.~C.}\ \bibnamefont
  {Newman}},\ }\href {\doibase 10.1088/0268-1242/2/5/013} {\bibfield  {journal}
  {\bibinfo  {journal} {Semiconductor Science and Technology}\ }\textbf
  {\bibinfo {volume} {2}},\ \bibinfo {pages} {315} (\bibinfo {year}
  {1987})}\BibitemShut {NoStop}%
\bibitem [{\citenamefont {Tersoff}(1990)}]{Tersoff1990}%
  \BibitemOpen
  \bibfield  {author} {\bibinfo {author} {\bibfnamefont {J.}~\bibnamefont
  {Tersoff}},\ }\href {\doibase 10.1103/physrevlett.64.1757} {\bibfield
  {journal} {\bibinfo  {journal} {Physical Review Letters}\ }\textbf {\bibinfo
  {volume} {64}},\ \bibinfo {pages} {1757} (\bibinfo {year}
  {1990})}\BibitemShut {NoStop}%
\bibitem [{\citenamefont {O'Donnell}\ \emph {et~al.}(1983)\citenamefont
  {O'Donnell}, \citenamefont {Lee},\ and\ \citenamefont
  {Watkins}}]{ODonnell1983}%
  \BibitemOpen
  \bibfield  {author} {\bibinfo {author} {\bibfnamefont {K.~P.}\ \bibnamefont
  {O'Donnell}}, \bibinfo {author} {\bibfnamefont {K.~M.}\ \bibnamefont {Lee}},
  \ and\ \bibinfo {author} {\bibfnamefont {G.~D.}\ \bibnamefont {Watkins}},\
  }\href {\doibase 10.1016/0378-4363(83)90256-5} {\bibfield  {journal}
  {\bibinfo  {journal} {Physica B+C}\ }\textbf {\bibinfo {volume} {116}},\
  \bibinfo {pages} {258} (\bibinfo {year} {1983})}\BibitemShut {NoStop}%
\bibitem [{\citenamefont {Londos}(1988)}]{Londos1988}%
  \BibitemOpen
  \bibfield  {author} {\bibinfo {author} {\bibfnamefont {C.~A.}\ \bibnamefont
  {Londos}},\ }\href {\doibase 10.1103/physrevb.37.4175} {\bibfield  {journal}
  {\bibinfo  {journal} {Physical Review B}\ }\textbf {\bibinfo {volume} {37}},\
  \bibinfo {pages} {4175} (\bibinfo {year} {1988})}\BibitemShut {NoStop}%
\bibitem [{\citenamefont {Trombetta}\ and\ \citenamefont
  {Watkins}(1987)}]{Trombetta1987}%
  \BibitemOpen
  \bibfield  {author} {\bibinfo {author} {\bibfnamefont {J.~M.}\ \bibnamefont
  {Trombetta}}\ and\ \bibinfo {author} {\bibfnamefont {G.~D.}\ \bibnamefont
  {Watkins}},\ }\href {\doibase 10.1063/1.98754} {\bibfield  {journal}
  {\bibinfo  {journal} {Applied Physics Letters}\ }\textbf {\bibinfo {volume}
  {51}},\ \bibinfo {pages} {1103} (\bibinfo {year} {1987})}\BibitemShut
  {NoStop}%
\bibitem [{\citenamefont {Asom}\ \emph {et~al.}(1987)\citenamefont {Asom},
  \citenamefont {Benton}, \citenamefont {Sauer},\ and\ \citenamefont
  {Kimerling}}]{Asom1987}%
  \BibitemOpen
  \bibfield  {author} {\bibinfo {author} {\bibfnamefont {M.~T.}\ \bibnamefont
  {Asom}}, \bibinfo {author} {\bibfnamefont {J.~L.}\ \bibnamefont {Benton}},
  \bibinfo {author} {\bibfnamefont {R.}~\bibnamefont {Sauer}}, \ and\ \bibinfo
  {author} {\bibfnamefont {L.~C.}\ \bibnamefont {Kimerling}},\ }\href {\doibase
  10.1063/1.98465} {\bibfield  {journal} {\bibinfo  {journal} {Applied Physics
  Letters}\ }\textbf {\bibinfo {volume} {51}},\ \bibinfo {pages} {256}
  (\bibinfo {year} {1987})}\BibitemShut {NoStop}%
\bibitem [{\citenamefont {Mooney}\ \emph {et~al.}(1977)\citenamefont {Mooney},
  \citenamefont {Cheng}, \citenamefont {S\"uli}, \citenamefont {Gerson},\ and\
  \citenamefont {Corbett}}]{Mooney1977}%
  \BibitemOpen
  \bibfield  {author} {\bibinfo {author} {\bibfnamefont {P.~M.}\ \bibnamefont
  {Mooney}}, \bibinfo {author} {\bibfnamefont {L.~J.}\ \bibnamefont {Cheng}},
  \bibinfo {author} {\bibfnamefont {M.}~\bibnamefont {S\"uli}}, \bibinfo
  {author} {\bibfnamefont {J.~D.}\ \bibnamefont {Gerson}}, \ and\ \bibinfo
  {author} {\bibfnamefont {J.~W.}\ \bibnamefont {Corbett}},\ }\href {\doibase
  10.1103/physrevb.15.3836} {\bibfield  {journal} {\bibinfo  {journal}
  {Physical Review B}\ }\textbf {\bibinfo {volume} {15}},\ \bibinfo {pages}
  {3836} (\bibinfo {year} {1977})}\BibitemShut {NoStop}%
\bibitem [{\citenamefont {Kimerling}\ \emph {et~al.}(1989)\citenamefont
  {Kimerling}, \citenamefont {Asom}, \citenamefont {Benton}, \citenamefont
  {Drevinsky},\ and\ \citenamefont {Caefer}}]{Kimerling1989}%
  \BibitemOpen
  \bibfield  {author} {\bibinfo {author} {\bibfnamefont {L.~C.}\ \bibnamefont
  {Kimerling}}, \bibinfo {author} {\bibfnamefont {M.~T.}\ \bibnamefont {Asom}},
  \bibinfo {author} {\bibfnamefont {J.~L.}\ \bibnamefont {Benton}}, \bibinfo
  {author} {\bibfnamefont {P.~J.}\ \bibnamefont {Drevinsky}}, \ and\ \bibinfo
  {author} {\bibfnamefont {C.~E.}\ \bibnamefont {Caefer}},\ }\href {\doibase
  10.4028/www.scientific.net/msf.38-41.141} {\bibfield  {journal} {\bibinfo
  {journal} {Materials Science Forum}\ }\textbf {\bibinfo {volume} {38-41}},\
  \bibinfo {pages} {141} (\bibinfo {year} {1989})}\BibitemShut {NoStop}%
\bibitem [{\citenamefont {Watkins}(1975)}]{Watkins1975}%
  \BibitemOpen
  \bibfield  {author} {\bibinfo {author} {\bibfnamefont {G.~D.}\ \bibnamefont
  {Watkins}},\ }\href {\doibase 10.1103/physrevb.12.5824} {\bibfield  {journal}
  {\bibinfo  {journal} {Physical Review B}\ }\textbf {\bibinfo {volume} {12}},\
  \bibinfo {pages} {5824} (\bibinfo {year} {1975})}\BibitemShut {NoStop}%
\bibitem [{\citenamefont {Vines}\ \emph {et~al.}(2008)\citenamefont {Vines},
  \citenamefont {Monakhov}, \citenamefont {Kuznetsov}, \citenamefont
  {Koz{\l}owski}, \citenamefont {Kaminski},\ and\ \citenamefont
  {Svensson}}]{Vines2008}%
  \BibitemOpen
  \bibfield  {author} {\bibinfo {author} {\bibfnamefont {L.}~\bibnamefont
  {Vines}}, \bibinfo {author} {\bibfnamefont {E.~V.}\ \bibnamefont {Monakhov}},
  \bibinfo {author} {\bibfnamefont {A.~Y.}\ \bibnamefont {Kuznetsov}}, \bibinfo
  {author} {\bibfnamefont {R.}~\bibnamefont {Koz{\l}owski}}, \bibinfo {author}
  {\bibfnamefont {P.}~\bibnamefont {Kaminski}}, \ and\ \bibinfo {author}
  {\bibfnamefont {B.~G.}\ \bibnamefont {Svensson}},\ }\href {\doibase
  10.1103/physrevb.78.085205} {\bibfield  {journal} {\bibinfo  {journal}
  {Physical Review B}\ }\textbf {\bibinfo {volume} {78}},\ \bibinfo {pages}
  {085205} (\bibinfo {year} {2008})}\BibitemShut {NoStop}%
\bibitem [{\citenamefont {Thonke}\ \emph {et~al.}(1984)\citenamefont {Thonke},
  \citenamefont {Watkins},\ and\ \citenamefont {Sauer}}]{Thonke1984}%
  \BibitemOpen
  \bibfield  {author} {\bibinfo {author} {\bibfnamefont {K.}~\bibnamefont
  {Thonke}}, \bibinfo {author} {\bibfnamefont {G.~D.}\ \bibnamefont {Watkins}},
  \ and\ \bibinfo {author} {\bibfnamefont {R.}~\bibnamefont {Sauer}},\ }\href
  {\doibase 10.1016/0038-1098(84)90532-5} {\bibfield  {journal} {\bibinfo
  {journal} {Solid State Communications}\ }\textbf {\bibinfo {volume} {51}},\
  \bibinfo {pages} {127} (\bibinfo {year} {1984})}\BibitemShut {NoStop}%
\bibitem [{\citenamefont {Davies}\ \emph {et~al.}(1986)\citenamefont {Davies},
  \citenamefont {Oates}, \citenamefont {Newman}, \citenamefont {Woolley},
  \citenamefont {Lightowlers}, \citenamefont {Binns},\ and\ \citenamefont
  {Wilkes}}]{Davies1986}%
  \BibitemOpen
  \bibfield  {author} {\bibinfo {author} {\bibfnamefont {G.}~\bibnamefont
  {Davies}}, \bibinfo {author} {\bibfnamefont {A.~S.}\ \bibnamefont {Oates}},
  \bibinfo {author} {\bibfnamefont {R.~C.}\ \bibnamefont {Newman}}, \bibinfo
  {author} {\bibfnamefont {R.}~\bibnamefont {Woolley}}, \bibinfo {author}
  {\bibfnamefont {E.~C.}\ \bibnamefont {Lightowlers}}, \bibinfo {author}
  {\bibfnamefont {M.~J.}\ \bibnamefont {Binns}}, \ and\ \bibinfo {author}
  {\bibfnamefont {J.~G.}\ \bibnamefont {Wilkes}},\ }\href {\doibase
  10.1088/0022-3719/19/6/006} {\bibfield  {journal} {\bibinfo  {journal}
  {Journal of Physics C: Solid State Physics}\ }\textbf {\bibinfo {volume}
  {19}},\ \bibinfo {pages} {841} (\bibinfo {year} {1986})}\BibitemShut
  {NoStop}%
\bibitem [{\citenamefont {K\"urner}\ \emph {et~al.}(1989)\citenamefont
  {K\"urner}, \citenamefont {Sauer}, \citenamefont {D\"ornen},\ and\
  \citenamefont {Thonke}}]{Kuerner1989}%
  \BibitemOpen
  \bibfield  {author} {\bibinfo {author} {\bibfnamefont {W.}~\bibnamefont
  {K\"urner}}, \bibinfo {author} {\bibfnamefont {R.}~\bibnamefont {Sauer}},
  \bibinfo {author} {\bibfnamefont {A.}~\bibnamefont {D\"ornen}}, \ and\
  \bibinfo {author} {\bibfnamefont {K.}~\bibnamefont {Thonke}},\ }\href
  {\doibase 10.1103/physrevb.39.13327} {\bibfield  {journal} {\bibinfo
  {journal} {Physical Review B}\ }\textbf {\bibinfo {volume} {39}},\ \bibinfo
  {pages} {13327} (\bibinfo {year} {1989})}\BibitemShut {NoStop}%
\bibitem [{\citenamefont {Svensson}\ and\ \citenamefont
  {Lindstr\"om}(1986{\natexlab{a}})}]{Svensson1986a}%
  \BibitemOpen
  \bibfield  {author} {\bibinfo {author} {\bibfnamefont {B.~G.}\ \bibnamefont
  {Svensson}}\ and\ \bibinfo {author} {\bibfnamefont {J.~L.}\ \bibnamefont
  {Lindstr\"om}},\ }\href {\doibase 10.1002/pssa.2210950222} {\bibfield
  {journal} {\bibinfo  {journal} {physica status solidi (a)}\ }\textbf
  {\bibinfo {volume} {95}},\ \bibinfo {pages} {537} (\bibinfo {year}
  {1986}{\natexlab{a}})}\BibitemShut {NoStop}%
\bibitem [{\citenamefont {Coutinho}\ \emph
  {et~al.}(2001{\natexlab{a}})\citenamefont {Coutinho}, \citenamefont {Jones},
  \citenamefont {Briddon}, \citenamefont {\"Oberg}, \citenamefont {Murin},
  \citenamefont {Markevich},\ and\ \citenamefont {Lindstr\"om}}]{Coutinho2001}%
  \BibitemOpen
  \bibfield  {author} {\bibinfo {author} {\bibfnamefont {J.}~\bibnamefont
  {Coutinho}}, \bibinfo {author} {\bibfnamefont {R.}~\bibnamefont {Jones}},
  \bibinfo {author} {\bibfnamefont {P.~R.}\ \bibnamefont {Briddon}}, \bibinfo
  {author} {\bibfnamefont {S.}~\bibnamefont {\"Oberg}}, \bibinfo {author}
  {\bibfnamefont {L.~I.}\ \bibnamefont {Murin}}, \bibinfo {author}
  {\bibfnamefont {V.~P.}\ \bibnamefont {Markevich}}, \ and\ \bibinfo {author}
  {\bibfnamefont {J.~L.}\ \bibnamefont {Lindstr\"om}},\ }\href {\doibase
  10.1103/physrevb.65.014109} {\bibfield  {journal} {\bibinfo  {journal}
  {Physical Review B}\ }\textbf {\bibinfo {volume} {65}},\ \bibinfo {pages}
  {014109} (\bibinfo {year} {2001}{\natexlab{a}})}\BibitemShut {NoStop}%
\bibitem [{\citenamefont {Murin}\ \emph {et~al.}(2001)\citenamefont {Murin},
  \citenamefont {Markevich}, \citenamefont {Lindstr\"om}, \citenamefont
  {Kleverman}, \citenamefont {Hermansson}, \citenamefont {Hallberg},\ and\
  \citenamefont {Svensson}}]{Murin2001}%
  \BibitemOpen
  \bibfield  {author} {\bibinfo {author} {\bibfnamefont {L.~I.}\ \bibnamefont
  {Murin}}, \bibinfo {author} {\bibfnamefont {V.~P.}\ \bibnamefont
  {Markevich}}, \bibinfo {author} {\bibfnamefont {J.~L.}\ \bibnamefont
  {Lindstr\"om}}, \bibinfo {author} {\bibfnamefont {M.}~\bibnamefont
  {Kleverman}}, \bibinfo {author} {\bibfnamefont {J.}~\bibnamefont
  {Hermansson}}, \bibinfo {author} {\bibfnamefont {T.}~\bibnamefont
  {Hallberg}}, \ and\ \bibinfo {author} {\bibfnamefont {B.~G.}\ \bibnamefont
  {Svensson}},\ }\href {\doibase 10.4028/www.scientific.net/ssp.82-84.57}
  {\bibfield  {journal} {\bibinfo  {journal} {Solid State Phenomena}\ }\textbf
  {\bibinfo {volume} {82-84}},\ \bibinfo {pages} {57} (\bibinfo {year}
  {2001})}\BibitemShut {NoStop}%
\bibitem [{\citenamefont {Jones}\ and\ \citenamefont
  {\"Oberg}(1992)}]{Jones1992}%
  \BibitemOpen
  \bibfield  {author} {\bibinfo {author} {\bibfnamefont {R.}~\bibnamefont
  {Jones}}\ and\ \bibinfo {author} {\bibfnamefont {S.}~\bibnamefont
  {\"Oberg}},\ }\href {\doibase 10.1103/physrevlett.68.86} {\bibfield
  {journal} {\bibinfo  {journal} {Physical Review Letters}\ }\textbf {\bibinfo
  {volume} {68}},\ \bibinfo {pages} {86} (\bibinfo {year} {1992})}\BibitemShut
  {NoStop}%
\bibitem [{\citenamefont {Khirunenko}\ \emph {et~al.}(2005)\citenamefont
  {Khirunenko}, \citenamefont {Pomozov}, \citenamefont {Tripachko},
  \citenamefont {Sosnin}, \citenamefont {Duvanskii}, \citenamefont {Murin},
  \citenamefont {Lindstr\"om}, \citenamefont {Lastovskii}, \citenamefont
  {Makarenko}, \citenamefont {Markevich},\ and\ \citenamefont
  {Peaker}}]{Khirunenko2005}%
  \BibitemOpen
  \bibfield  {author} {\bibinfo {author} {\bibfnamefont {L.~I.}\ \bibnamefont
  {Khirunenko}}, \bibinfo {author} {\bibfnamefont {Y.~V.}\ \bibnamefont
  {Pomozov}}, \bibinfo {author} {\bibfnamefont {N.~A.}\ \bibnamefont
  {Tripachko}}, \bibinfo {author} {\bibfnamefont {M.~G.}\ \bibnamefont
  {Sosnin}}, \bibinfo {author} {\bibfnamefont {A.~V.}\ \bibnamefont
  {Duvanskii}}, \bibinfo {author} {\bibfnamefont {L.~I.}\ \bibnamefont
  {Murin}}, \bibinfo {author} {\bibfnamefont {J.~L.}\ \bibnamefont
  {Lindstr\"om}}, \bibinfo {author} {\bibfnamefont {S.~B.}\ \bibnamefont
  {Lastovskii}}, \bibinfo {author} {\bibfnamefont {L.~F.}\ \bibnamefont
  {Makarenko}}, \bibinfo {author} {\bibfnamefont {V.~P.}\ \bibnamefont
  {Markevich}}, \ and\ \bibinfo {author} {\bibfnamefont {A.~R.}\ \bibnamefont
  {Peaker}},\ }\href {\doibase 10.4028/www.scientific.net/ssp.108-109.261}
  {\bibfield  {journal} {\bibinfo  {journal} {Solid State Phenomena}\ }\textbf
  {\bibinfo {volume} {108-109}},\ \bibinfo {pages} {261} (\bibinfo {year}
  {2005})}\BibitemShut {NoStop}%
\bibitem [{\citenamefont {Khirunenko}\ \emph {et~al.}(2008)\citenamefont
  {Khirunenko}, \citenamefont {Sosnin}, \citenamefont {Pomozov}, \citenamefont
  {Murin}, \citenamefont {Markevich}, \citenamefont {Peaker}, \citenamefont
  {Almeida}, \citenamefont {Coutinho},\ and\ \citenamefont
  {Torres}}]{Khirunenko2008}%
  \BibitemOpen
  \bibfield  {author} {\bibinfo {author} {\bibfnamefont {L.~I.}\ \bibnamefont
  {Khirunenko}}, \bibinfo {author} {\bibfnamefont {M.~G.}\ \bibnamefont
  {Sosnin}}, \bibinfo {author} {\bibfnamefont {Y.~V.}\ \bibnamefont {Pomozov}},
  \bibinfo {author} {\bibfnamefont {L.~I.}\ \bibnamefont {Murin}}, \bibinfo
  {author} {\bibfnamefont {V.~P.}\ \bibnamefont {Markevich}}, \bibinfo {author}
  {\bibfnamefont {A.~R.}\ \bibnamefont {Peaker}}, \bibinfo {author}
  {\bibfnamefont {L.~M.}\ \bibnamefont {Almeida}}, \bibinfo {author}
  {\bibfnamefont {J.}~\bibnamefont {Coutinho}}, \ and\ \bibinfo {author}
  {\bibfnamefont {V.~J.~B.}\ \bibnamefont {Torres}},\ }\href {\doibase
  10.1103/physrevb.78.155203} {\bibfield  {journal} {\bibinfo  {journal}
  {Physical Review B}\ }\textbf {\bibinfo {volume} {78}},\ \bibinfo {pages}
  {155203} (\bibinfo {year} {2008})}\BibitemShut {NoStop}%
\bibitem [{\citenamefont {Abdullin}\ \emph {et~al.}(1990)\citenamefont
  {Abdullin}, \citenamefont {Mukashev}, \citenamefont {Tamendarov},\ and\
  \citenamefont {Tashenov}}]{Abdullin1990}%
  \BibitemOpen
  \bibfield  {author} {\bibinfo {author} {\bibfnamefont {K.~A.}\ \bibnamefont
  {Abdullin}}, \bibinfo {author} {\bibfnamefont {B.~N.}\ \bibnamefont
  {Mukashev}}, \bibinfo {author} {\bibfnamefont {M.~F.}\ \bibnamefont
  {Tamendarov}}, \ and\ \bibinfo {author} {\bibfnamefont {T.~B.}\ \bibnamefont
  {Tashenov}},\ }\href {\doibase 10.1016/0375-9601(90)90700-x} {\bibfield
  {journal} {\bibinfo  {journal} {Physics Letters A}\ }\textbf {\bibinfo
  {volume} {144}},\ \bibinfo {pages} {198} (\bibinfo {year}
  {1990})}\BibitemShut {NoStop}%
\bibitem [{\citenamefont {Davies}\ and\ \citenamefont
  {Newman}(1994)}]{Davies1994}%
  \BibitemOpen
  \bibfield  {author} {\bibinfo {author} {\bibfnamefont {G.}~\bibnamefont
  {Davies}}\ and\ \bibinfo {author} {\bibfnamefont {R.~C.}\ \bibnamefont
  {Newman}},\ }\enquote {\bibinfo {title} {Carbon in monocrystalline
  silicon},}\ in\ \href@noop {} {\emph {\bibinfo {booktitle} {Handbook on
  Semiconductors}}},\ Vol.~\bibinfo {volume} {3}\ (\bibinfo  {publisher}
  {Mahajan, S., Ed., Amsterdam: Elsevier,},\ \bibinfo {year} {1994})\ p.\
  \bibinfo {pages} {1557}\BibitemShut {NoStop}%
\bibitem [{\citenamefont {Davies}\ \emph {et~al.}(2006)\citenamefont {Davies},
  \citenamefont {Hayama}, \citenamefont {Murin}, \citenamefont
  {Krause-Rehberg}, \citenamefont {Bondarenko}, \citenamefont {Sengupta},
  \citenamefont {Davia},\ and\ \citenamefont {Karpenko}}]{Davies2006}%
  \BibitemOpen
  \bibfield  {author} {\bibinfo {author} {\bibfnamefont {G.}~\bibnamefont
  {Davies}}, \bibinfo {author} {\bibfnamefont {S.}~\bibnamefont {Hayama}},
  \bibinfo {author} {\bibfnamefont {L.}~\bibnamefont {Murin}}, \bibinfo
  {author} {\bibfnamefont {R.}~\bibnamefont {Krause-Rehberg}}, \bibinfo
  {author} {\bibfnamefont {V.}~\bibnamefont {Bondarenko}}, \bibinfo {author}
  {\bibfnamefont {A.}~\bibnamefont {Sengupta}}, \bibinfo {author}
  {\bibfnamefont {C.}~\bibnamefont {Davia}}, \ and\ \bibinfo {author}
  {\bibfnamefont {A.}~\bibnamefont {Karpenko}},\ }\href {\doibase
  10.1103/physrevb.73.165202} {\bibfield  {journal} {\bibinfo  {journal}
  {Physical Review B}\ }\textbf {\bibinfo {volume} {73}},\ \bibinfo {pages}
  {165202} (\bibinfo {year} {2006})}\BibitemShut {NoStop}%
\bibitem [{\citenamefont {Fukuoka}\ \emph {et~al.}(1990)\citenamefont
  {Fukuoka}, \citenamefont {Atobe},\ and\ \citenamefont {Honda}}]{Fukuoka1990}%
  \BibitemOpen
  \bibfield  {author} {\bibinfo {author} {\bibfnamefont {N.}~\bibnamefont
  {Fukuoka}}, \bibinfo {author} {\bibfnamefont {K.}~\bibnamefont {Atobe}}, \
  and\ \bibinfo {author} {\bibfnamefont {M.}~\bibnamefont {Honda}},\ }\href
  {\doibase 10.1143/jjap.29.1625} {\bibfield  {journal} {\bibinfo  {journal}
  {Japanese Journal of Applied Physics}\ }\textbf {\bibinfo {volume} {29}},\
  \bibinfo {pages} {1625} (\bibinfo {year} {1990})}\BibitemShut {NoStop}%
\bibitem [{\citenamefont {Awadelkarim}\ \emph {et~al.}(1990)\citenamefont
  {Awadelkarim}, \citenamefont {Henry}, \citenamefont {Monemar}, \citenamefont
  {Lindstr\"om}, \citenamefont {Zhang},\ and\ \citenamefont
  {Corbett}}]{Awadelkarim1990}%
  \BibitemOpen
  \bibfield  {author} {\bibinfo {author} {\bibfnamefont {O.~O.}\ \bibnamefont
  {Awadelkarim}}, \bibinfo {author} {\bibfnamefont {A.}~\bibnamefont {Henry}},
  \bibinfo {author} {\bibfnamefont {B.}~\bibnamefont {Monemar}}, \bibinfo
  {author} {\bibfnamefont {J.~L.}\ \bibnamefont {Lindstr\"om}}, \bibinfo
  {author} {\bibfnamefont {Y.}~\bibnamefont {Zhang}}, \ and\ \bibinfo {author}
  {\bibfnamefont {J.~W.}\ \bibnamefont {Corbett}},\ }\href {\doibase
  10.1103/physrevb.42.5635} {\bibfield  {journal} {\bibinfo  {journal}
  {Physical Review B}\ }\textbf {\bibinfo {volume} {42}},\ \bibinfo {pages}
  {5635} (\bibinfo {year} {1990})}\BibitemShut {NoStop}%
\bibitem [{\citenamefont {Ganagona}\ \emph {et~al.}(2012)\citenamefont
  {Ganagona}, \citenamefont {Raeissi}, \citenamefont {Vines}, \citenamefont
  {Monakhov},\ and\ \citenamefont {Svensson}}]{Ganagona2012}%
  \BibitemOpen
  \bibfield  {author} {\bibinfo {author} {\bibfnamefont {N.}~\bibnamefont
  {Ganagona}}, \bibinfo {author} {\bibfnamefont {B.}~\bibnamefont {Raeissi}},
  \bibinfo {author} {\bibfnamefont {L.}~\bibnamefont {Vines}}, \bibinfo
  {author} {\bibfnamefont {E.~V.}\ \bibnamefont {Monakhov}}, \ and\ \bibinfo
  {author} {\bibfnamefont {B.~G.}\ \bibnamefont {Svensson}},\ }\href {\doibase
  10.1002/pssc.201200217} {\bibfield  {journal} {\bibinfo  {journal} {physica
  status solidi (c)}\ }\textbf {\bibinfo {volume} {9}},\ \bibinfo {pages}
  {2009} (\bibinfo {year} {2012})}\BibitemShut {NoStop}%
\bibitem [{\citenamefont {Raeissi}\ \emph {et~al.}(2013)\citenamefont
  {Raeissi}, \citenamefont {Ganagona}, \citenamefont {Galeckas}, \citenamefont
  {Monakhov},\ and\ \citenamefont {Svensson}}]{Raeissi2013}%
  \BibitemOpen
  \bibfield  {author} {\bibinfo {author} {\bibfnamefont {B.}~\bibnamefont
  {Raeissi}}, \bibinfo {author} {\bibfnamefont {N.}~\bibnamefont {Ganagona}},
  \bibinfo {author} {\bibfnamefont {A.}~\bibnamefont {Galeckas}}, \bibinfo
  {author} {\bibfnamefont {E.~V.}\ \bibnamefont {Monakhov}}, \ and\ \bibinfo
  {author} {\bibfnamefont {B.~G.}\ \bibnamefont {Svensson}},\ }\href {\doibase
  10.4028/www.scientific.net/ssp.205-206.224} {\bibfield  {journal} {\bibinfo
  {journal} {Solid State Phenomena}\ }\textbf {\bibinfo {volume} {205-206}},\
  \bibinfo {pages} {224} (\bibinfo {year} {2013})}\BibitemShut {NoStop}%
\bibitem [{\citenamefont {Ayedh}\ \emph {et~al.}(2019)\citenamefont {Ayedh},
  \citenamefont {Grigorev}, \citenamefont {Galeckas}, \citenamefont
  {Svensson},\ and\ \citenamefont {Monakhov}}]{Ayedh2019}%
  \BibitemOpen
  \bibfield  {author} {\bibinfo {author} {\bibfnamefont {H.~M.}\ \bibnamefont
  {Ayedh}}, \bibinfo {author} {\bibfnamefont {A.~A.}\ \bibnamefont {Grigorev}},
  \bibinfo {author} {\bibfnamefont {A.}~\bibnamefont {Galeckas}}, \bibinfo
  {author} {\bibfnamefont {B.~G.}\ \bibnamefont {Svensson}}, \ and\ \bibinfo
  {author} {\bibfnamefont {E.~V.}\ \bibnamefont {Monakhov}},\ }\href {\doibase
  10.1002/pssa.201800986} {\bibfield  {journal} {\bibinfo  {journal} {Physica
  Status Solidi A}\ }\textbf {\bibinfo {volume} {216}},\ \bibinfo {pages}
  {1800986} (\bibinfo {year} {2019})}\BibitemShut {NoStop}%
\bibitem [{\citenamefont {Kresse}\ and\ \citenamefont
  {Hafner}(1993)}]{Kresse1993}%
  \BibitemOpen
  \bibfield  {author} {\bibinfo {author} {\bibfnamefont {G.}~\bibnamefont
  {Kresse}}\ and\ \bibinfo {author} {\bibfnamefont {J.}~\bibnamefont
  {Hafner}},\ }\href {\doibase 10.1103/PhysRevB.47.558} {\bibfield  {journal}
  {\bibinfo  {journal} {Physical Review B}\ }\textbf {\bibinfo {volume} {47}},\
  \bibinfo {pages} {558} (\bibinfo {year} {1993})}\BibitemShut {NoStop}%
\bibitem [{\citenamefont {Kresse}\ and\ \citenamefont
  {Hafner}(1994)}]{Kresse1994}%
  \BibitemOpen
  \bibfield  {author} {\bibinfo {author} {\bibfnamefont {G.}~\bibnamefont
  {Kresse}}\ and\ \bibinfo {author} {\bibfnamefont {J.}~\bibnamefont
  {Hafner}},\ }\href {\doibase 10.1103/PhysRevB.49.14251} {\bibfield  {journal}
  {\bibinfo  {journal} {Physical Review B}\ }\textbf {\bibinfo {volume} {49}},\
  \bibinfo {pages} {14251} (\bibinfo {year} {1994})}\BibitemShut {NoStop}%
\bibitem [{\citenamefont {Kresse}\ and\ \citenamefont
  {Furthm{\"u}ller}(1996{\natexlab{a}})}]{Kresse1996}%
  \BibitemOpen
  \bibfield  {author} {\bibinfo {author} {\bibfnamefont {G.}~\bibnamefont
  {Kresse}}\ and\ \bibinfo {author} {\bibfnamefont {J.}~\bibnamefont
  {Furthm{\"u}ller}},\ }\href {\doibase 10.1103/PhysRevB.54.11169} {\bibfield
  {journal} {\bibinfo  {journal} {Physical Review B}\ }\textbf {\bibinfo
  {volume} {54}},\ \bibinfo {pages} {11169} (\bibinfo {year}
  {1996}{\natexlab{a}})}\BibitemShut {NoStop}%
\bibitem [{\citenamefont {Kresse}\ and\ \citenamefont
  {Furthm{\"u}ller}(1996{\natexlab{b}})}]{Kresse1996a}%
  \BibitemOpen
  \bibfield  {author} {\bibinfo {author} {\bibfnamefont {G.}~\bibnamefont
  {Kresse}}\ and\ \bibinfo {author} {\bibfnamefont {J.}~\bibnamefont
  {Furthm{\"u}ller}},\ }\href {\doibase 10.1016/0927-0256(96)00008-0}
  {\bibfield  {journal} {\bibinfo  {journal} {Computational Materials Science}\
  }\textbf {\bibinfo {volume} {6}},\ \bibinfo {pages} {15} (\bibinfo {year}
  {1996}{\natexlab{b}})}\BibitemShut {NoStop}%
\bibitem [{\citenamefont {Bl\"{o}chl}(1994)}]{Blochl1994}%
  \BibitemOpen
  \bibfield  {author} {\bibinfo {author} {\bibfnamefont {P.~E.}\ \bibnamefont
  {Bl\"{o}chl}},\ }\href {\doibase 10.1103/physrevb.50.17953} {\bibfield
  {journal} {\bibinfo  {journal} {Physical Review B}\ }\textbf {\bibinfo
  {volume} {50}},\ \bibinfo {pages} {17953} (\bibinfo {year}
  {1994})}\BibitemShut {NoStop}%
\bibitem [{\citenamefont {Heyd}\ \emph {et~al.}(2003)\citenamefont {Heyd},
  \citenamefont {Scuseria},\ and\ \citenamefont {Ernzerhof}}]{Heyd2003}%
  \BibitemOpen
  \bibfield  {author} {\bibinfo {author} {\bibfnamefont {J.}~\bibnamefont
  {Heyd}}, \bibinfo {author} {\bibfnamefont {G.~E.}\ \bibnamefont {Scuseria}},
  \ and\ \bibinfo {author} {\bibfnamefont {M.}~\bibnamefont {Ernzerhof}},\
  }\href {\doibase 10.1063/1.1564060} {\bibfield  {journal} {\bibinfo
  {journal} {Journal of Chemical Physics}\ }\textbf {\bibinfo {volume} {118}},\
  \bibinfo {pages} {8207} (\bibinfo {year} {2003})}\BibitemShut {NoStop}%
\bibitem [{\citenamefont {Krukau}\ \emph {et~al.}(2006)\citenamefont {Krukau},
  \citenamefont {Vydrov}, \citenamefont {Izmaylov},\ and\ \citenamefont
  {Scuseria}}]{Krukau2006}%
  \BibitemOpen
  \bibfield  {author} {\bibinfo {author} {\bibfnamefont {A.~V.}\ \bibnamefont
  {Krukau}}, \bibinfo {author} {\bibfnamefont {O.~A.}\ \bibnamefont {Vydrov}},
  \bibinfo {author} {\bibfnamefont {A.~F.}\ \bibnamefont {Izmaylov}}, \ and\
  \bibinfo {author} {\bibfnamefont {G.~E.}\ \bibnamefont {Scuseria}},\ }\href
  {\doibase 10.1063/1.2404663} {\bibfield  {journal} {\bibinfo  {journal}
  {Journal of Chemical Physics}\ }\textbf {\bibinfo {volume} {125}},\ \bibinfo
  {pages} {224106} (\bibinfo {year} {2006})}\BibitemShut {NoStop}%
\bibitem [{\citenamefont {Perdew}\ \emph {et~al.}(1996)\citenamefont {Perdew},
  \citenamefont {Burke},\ and\ \citenamefont {Ernzerhof}}]{Perdew1996}%
  \BibitemOpen
  \bibfield  {author} {\bibinfo {author} {\bibfnamefont {J.~P.}\ \bibnamefont
  {Perdew}}, \bibinfo {author} {\bibfnamefont {K.}~\bibnamefont {Burke}}, \
  and\ \bibinfo {author} {\bibfnamefont {M.}~\bibnamefont {Ernzerhof}},\ }\href
  {\doibase 10.1103/physrevlett.77.3865} {\bibfield  {journal} {\bibinfo
  {journal} {Physical Review Letters}\ }\textbf {\bibinfo {volume} {77}},\
  \bibinfo {pages} {3865} (\bibinfo {year} {1996})}\BibitemShut {NoStop}%
\bibitem [{\citenamefont {Gouveia}\ and\ \citenamefont
  {Coutinho}(2019)}]{Gouveia2019}%
  \BibitemOpen
  \bibfield  {author} {\bibinfo {author} {\bibfnamefont {J.~D.}\ \bibnamefont
  {Gouveia}}\ and\ \bibinfo {author} {\bibfnamefont {J.}~\bibnamefont
  {Coutinho}},\ }\href {\doibase 10.1088/2516-1075/aafc4b} {\bibfield
  {journal} {\bibinfo  {journal} {Electronic Structure}\ }\textbf {\bibinfo
  {volume} {1}},\ \bibinfo {pages} {015008} (\bibinfo {year}
  {2019})}\BibitemShut {NoStop}%
\bibitem [{\citenamefont {Bathen}\ \emph {et~al.}(2019)\citenamefont {Bathen},
  \citenamefont {Coutinho}, \citenamefont {Ayedh}, \citenamefont {Hassan},
  \citenamefont {Farkas}, \citenamefont {Öberg}, \citenamefont {Frodason},
  \citenamefont {Svensson},\ and\ \citenamefont {Vines}}]{Bathen2019}%
  \BibitemOpen
  \bibfield  {author} {\bibinfo {author} {\bibfnamefont {M.~E.}\ \bibnamefont
  {Bathen}}, \bibinfo {author} {\bibfnamefont {J.}~\bibnamefont {Coutinho}},
  \bibinfo {author} {\bibfnamefont {H.~M.}\ \bibnamefont {Ayedh}}, \bibinfo
  {author} {\bibfnamefont {J.~U.}\ \bibnamefont {Hassan}}, \bibinfo {author}
  {\bibfnamefont {I.}~\bibnamefont {Farkas}}, \bibinfo {author} {\bibfnamefont
  {S.}~\bibnamefont {Öberg}}, \bibinfo {author} {\bibfnamefont {Y.~K.}\
  \bibnamefont {Frodason}}, \bibinfo {author} {\bibfnamefont {B.~G.}\
  \bibnamefont {Svensson}}, \ and\ \bibinfo {author} {\bibfnamefont
  {L.}~\bibnamefont {Vines}},\ }\href {\doibase 10.1103/physrevb.100.014103}
  {\bibfield  {journal} {\bibinfo  {journal} {Physical Review B}\ }\textbf
  {\bibinfo {volume} {100}},\ \bibinfo {pages} {014103} (\bibinfo {year}
  {2019})}\BibitemShut {NoStop}%
\bibitem [{\citenamefont {Mills}\ and\ \citenamefont
  {J{\'{o}}nsson}(1994)}]{Mills1994}%
  \BibitemOpen
  \bibfield  {author} {\bibinfo {author} {\bibfnamefont {G.}~\bibnamefont
  {Mills}}\ and\ \bibinfo {author} {\bibfnamefont {H.}~\bibnamefont
  {J{\'{o}}nsson}},\ }\href {\doibase 10.1103/physrevlett.72.1124} {\bibfield
  {journal} {\bibinfo  {journal} {Physical Review Letters}\ }\textbf {\bibinfo
  {volume} {72}},\ \bibinfo {pages} {1124} (\bibinfo {year}
  {1994})}\BibitemShut {NoStop}%
\bibitem [{\citenamefont {Mills}\ \emph {et~al.}(1995)\citenamefont {Mills},
  \citenamefont {J{\'{o}}nsson},\ and\ \citenamefont {Schenter}}]{Mills1995}%
  \BibitemOpen
  \bibfield  {author} {\bibinfo {author} {\bibfnamefont {G.}~\bibnamefont
  {Mills}}, \bibinfo {author} {\bibfnamefont {H.}~\bibnamefont
  {J{\'{o}}nsson}}, \ and\ \bibinfo {author} {\bibfnamefont {G.~K.}\
  \bibnamefont {Schenter}},\ }\href {\doibase 10.1016/0039-6028(94)00731-4}
  {\bibfield  {journal} {\bibinfo  {journal} {Surface Science}\ }\textbf
  {\bibinfo {volume} {324}},\ \bibinfo {pages} {305} (\bibinfo {year}
  {1995})}\BibitemShut {NoStop}%
\bibitem [{\citenamefont {Henkelman}\ and\ \citenamefont
  {J{\'{o}}nsson}(1999)}]{Henkelman1999}%
  \BibitemOpen
  \bibfield  {author} {\bibinfo {author} {\bibfnamefont {G.}~\bibnamefont
  {Henkelman}}\ and\ \bibinfo {author} {\bibfnamefont {H.}~\bibnamefont
  {J{\'{o}}nsson}},\ }\href {\doibase 10.1063/1.480097} {\bibfield  {journal}
  {\bibinfo  {journal} {Journal of Chemical Physics}\ }\textbf {\bibinfo
  {volume} {111}},\ \bibinfo {pages} {7010} (\bibinfo {year}
  {1999})}\BibitemShut {NoStop}%
\bibitem [{\citenamefont {Henkelman}\ \emph {et~al.}(2000)\citenamefont
  {Henkelman}, \citenamefont {Uberuaga},\ and\ \citenamefont
  {J{\'{o}}nsson}}]{Henkelman2000}%
  \BibitemOpen
  \bibfield  {author} {\bibinfo {author} {\bibfnamefont {G.}~\bibnamefont
  {Henkelman}}, \bibinfo {author} {\bibfnamefont {B.~P.}\ \bibnamefont
  {Uberuaga}}, \ and\ \bibinfo {author} {\bibfnamefont {H.}~\bibnamefont
  {J{\'{o}}nsson}},\ }\href {\doibase 10.1063/1.1329672} {\bibfield  {journal}
  {\bibinfo  {journal} {Journal of Chemical Physics}\ }\textbf {\bibinfo
  {volume} {113}},\ \bibinfo {pages} {9901} (\bibinfo {year}
  {2000})}\BibitemShut {NoStop}%
\bibitem [{\citenamefont {Svensson}\ \emph {et~al.}(1989)\citenamefont
  {Svensson}, \citenamefont {Ryd{\'{e}}n},\ and\ \citenamefont
  {Lewerentz}}]{Svensson1989}%
  \BibitemOpen
  \bibfield  {author} {\bibinfo {author} {\bibfnamefont {B.~G.}\ \bibnamefont
  {Svensson}}, \bibinfo {author} {\bibfnamefont {K.-H.}\ \bibnamefont
  {Ryd{\'{e}}n}}, \ and\ \bibinfo {author} {\bibfnamefont {B.~M.~S.}\
  \bibnamefont {Lewerentz}},\ }\href {\doibase 10.1063/1.344389} {\bibfield
  {journal} {\bibinfo  {journal} {Journal of Applied Physics}\ }\textbf
  {\bibinfo {volume} {66}},\ \bibinfo {pages} {1699} (\bibinfo {year}
  {1989})}\BibitemShut {NoStop}%
\bibitem [{\citenamefont {Istratov}(1997)}]{Istratov1997}%
  \BibitemOpen
  \bibfield  {author} {\bibinfo {author} {\bibfnamefont {A.~A.}\ \bibnamefont
  {Istratov}},\ }\href {\doibase 10.1063/1.366269} {\bibfield  {journal}
  {\bibinfo  {journal} {Journal of Applied Physics}\ }\textbf {\bibinfo
  {volume} {82}},\ \bibinfo {pages} {2965} (\bibinfo {year}
  {1997})}\BibitemShut {NoStop}%
\bibitem [{\citenamefont {Song}\ and\ \citenamefont
  {Watkins}(1990)}]{Song1990}%
  \BibitemOpen
  \bibfield  {author} {\bibinfo {author} {\bibfnamefont {L.~W.}\ \bibnamefont
  {Song}}\ and\ \bibinfo {author} {\bibfnamefont {G.~D.}\ \bibnamefont
  {Watkins}},\ }\href {\doibase 10.1103/physrevb.42.5759} {\bibfield  {journal}
  {\bibinfo  {journal} {Physical Review B}\ }\textbf {\bibinfo {volume} {42}},\
  \bibinfo {pages} {5759} (\bibinfo {year} {1990})}\BibitemShut {NoStop}%
\bibitem [{\citenamefont {Bosomworth}\ \emph {et~al.}(1970)\citenamefont
  {Bosomworth}, \citenamefont {Hayes}, \citenamefont {Spray},\ and\
  \citenamefont {Watkins}}]{Bosomworth1970}%
  \BibitemOpen
  \bibfield  {author} {\bibinfo {author} {\bibfnamefont {D.~R.}\ \bibnamefont
  {Bosomworth}}, \bibinfo {author} {\bibfnamefont {W.}~\bibnamefont {Hayes}},
  \bibinfo {author} {\bibfnamefont {A.~R.~L.}\ \bibnamefont {Spray}}, \ and\
  \bibinfo {author} {\bibfnamefont {G.~D.}\ \bibnamefont {Watkins}},\ }\href
  {\doibase 10.1098/rspa.1970.0107} {\bibfield  {journal} {\bibinfo  {journal}
  {Proceedings of the Royal Society A. Mathematical, Physical and Engineering
  Sciences}\ }\textbf {\bibinfo {volume} {317}},\ \bibinfo {pages} {133}
  (\bibinfo {year} {1970})}\BibitemShut {NoStop}%
\bibitem [{\citenamefont {Pesola}\ \emph {et~al.}(1999)\citenamefont {Pesola},
  \citenamefont {von Boehm}, \citenamefont {Mattila},\ and\ \citenamefont
  {Nieminen}}]{Pesola1999}%
  \BibitemOpen
  \bibfield  {author} {\bibinfo {author} {\bibfnamefont {M.}~\bibnamefont
  {Pesola}}, \bibinfo {author} {\bibfnamefont {J.}~\bibnamefont {von Boehm}},
  \bibinfo {author} {\bibfnamefont {T.}~\bibnamefont {Mattila}}, \ and\
  \bibinfo {author} {\bibfnamefont {R.~M.}\ \bibnamefont {Nieminen}},\ }\href
  {\doibase 10.1103/physrevb.60.11449} {\bibfield  {journal} {\bibinfo
  {journal} {Physical Review B}\ }\textbf {\bibinfo {volume} {60}},\ \bibinfo
  {pages} {11449} (\bibinfo {year} {1999})}\BibitemShut {NoStop}%
\bibitem [{\citenamefont {Coutinho}\ \emph {et~al.}(2000)\citenamefont
  {Coutinho}, \citenamefont {Jones}, \citenamefont {Briddon},\ and\
  \citenamefont {\"Oberg}}]{Coutinho2000}%
  \BibitemOpen
  \bibfield  {author} {\bibinfo {author} {\bibfnamefont {J.}~\bibnamefont
  {Coutinho}}, \bibinfo {author} {\bibfnamefont {R.}~\bibnamefont {Jones}},
  \bibinfo {author} {\bibfnamefont {P.~R.}\ \bibnamefont {Briddon}}, \ and\
  \bibinfo {author} {\bibfnamefont {S.}~\bibnamefont {\"Oberg}},\ }\href
  {\doibase 10.1103/physrevb.62.10824} {\bibfield  {journal} {\bibinfo
  {journal} {Physical Review B}\ }\textbf {\bibinfo {volume} {62}},\ \bibinfo
  {pages} {10824} (\bibinfo {year} {2000})}\BibitemShut {NoStop}%
\bibitem [{\citenamefont {Needels}\ \emph {et~al.}(1991)\citenamefont
  {Needels}, \citenamefont {Joannopoulos}, \citenamefont {Bar-Yam},\ and\
  \citenamefont {Pantelides}}]{Needels1991}%
  \BibitemOpen
  \bibfield  {author} {\bibinfo {author} {\bibfnamefont {M.}~\bibnamefont
  {Needels}}, \bibinfo {author} {\bibfnamefont {J.~D.}\ \bibnamefont
  {Joannopoulos}}, \bibinfo {author} {\bibfnamefont {Y.}~\bibnamefont
  {Bar-Yam}}, \ and\ \bibinfo {author} {\bibfnamefont {S.~T.}\ \bibnamefont
  {Pantelides}},\ }\href {\doibase 10.1103/physrevb.43.4208} {\bibfield
  {journal} {\bibinfo  {journal} {Physical Review B}\ }\textbf {\bibinfo
  {volume} {43}},\ \bibinfo {pages} {4208} (\bibinfo {year}
  {1991})}\BibitemShut {NoStop}%
\bibitem [{\citenamefont {Lastovskii}\ \emph {et~al.}(2017)\citenamefont
  {Lastovskii}, \citenamefont {Gusakov}, \citenamefont {Markevich},
  \citenamefont {Peaker}, \citenamefont {Yakushevich}, \citenamefont
  {Korshunov},\ and\ \citenamefont {Murin}}]{Lastovskii2017}%
  \BibitemOpen
  \bibfield  {author} {\bibinfo {author} {\bibfnamefont {S.~B.}\ \bibnamefont
  {Lastovskii}}, \bibinfo {author} {\bibfnamefont {V.~E.}\ \bibnamefont
  {Gusakov}}, \bibinfo {author} {\bibfnamefont {V.~P.}\ \bibnamefont
  {Markevich}}, \bibinfo {author} {\bibfnamefont {A.~R.}\ \bibnamefont
  {Peaker}}, \bibinfo {author} {\bibfnamefont {H.~S.}\ \bibnamefont
  {Yakushevich}}, \bibinfo {author} {\bibfnamefont {F.~P.}\ \bibnamefont
  {Korshunov}}, \ and\ \bibinfo {author} {\bibfnamefont {L.~I.}\ \bibnamefont
  {Murin}},\ }\href {\doibase 10.1002/pssa.201700262} {\bibfield  {journal}
  {\bibinfo  {journal} {physica status solidi (a)}\ }\textbf {\bibinfo {volume}
  {214}},\ \bibinfo {pages} {1700262} (\bibinfo {year} {2017})}\BibitemShut
  {NoStop}%
\bibitem [{\citenamefont {Corbett}\ \emph {et~al.}(1964)\citenamefont
  {Corbett}, \citenamefont {McDonald},\ and\ \citenamefont
  {Watkins}}]{Corbett1964}%
  \BibitemOpen
  \bibfield  {author} {\bibinfo {author} {\bibfnamefont {J.~W.}\ \bibnamefont
  {Corbett}}, \bibinfo {author} {\bibfnamefont {R.~S.}\ \bibnamefont
  {McDonald}}, \ and\ \bibinfo {author} {\bibfnamefont {G.~D.}\ \bibnamefont
  {Watkins}},\ }\href {\doibase 10.1016/0022-3697(64)90100-3} {\bibfield
  {journal} {\bibinfo  {journal} {Journal of Physics and Chemistry of Solids}\
  }\textbf {\bibinfo {volume} {25}},\ \bibinfo {pages} {873} (\bibinfo {year}
  {1964})}\BibitemShut {NoStop}%
\bibitem [{\citenamefont {Quemener}\ \emph {et~al.}(2015)\citenamefont
  {Quemener}, \citenamefont {Raeissi}, \citenamefont {Herklotz}, \citenamefont
  {Murin}, \citenamefont {Monakhov},\ and\ \citenamefont
  {Svensson}}]{Quemener2015}%
  \BibitemOpen
  \bibfield  {author} {\bibinfo {author} {\bibfnamefont {V.}~\bibnamefont
  {Quemener}}, \bibinfo {author} {\bibfnamefont {B.}~\bibnamefont {Raeissi}},
  \bibinfo {author} {\bibfnamefont {F.}~\bibnamefont {Herklotz}}, \bibinfo
  {author} {\bibfnamefont {L.~I.}\ \bibnamefont {Murin}}, \bibinfo {author}
  {\bibfnamefont {E.~V.}\ \bibnamefont {Monakhov}}, \ and\ \bibinfo {author}
  {\bibfnamefont {B.~G.}\ \bibnamefont {Svensson}},\ }\href {\doibase
  10.1063/1.4932019} {\bibfield  {journal} {\bibinfo  {journal} {Journal of
  Applied Physics}\ }\textbf {\bibinfo {volume} {118}},\ \bibinfo {pages}
  {135703} (\bibinfo {year} {2015})}\BibitemShut {NoStop}%
\bibitem [{\citenamefont {Capaz}\ \emph {et~al.}(1994)\citenamefont {Capaz},
  \citenamefont {{Dal Pino}},\ and\ \citenamefont {Joannopoulos}}]{Capaz1994}%
  \BibitemOpen
  \bibfield  {author} {\bibinfo {author} {\bibfnamefont {R.~B.}\ \bibnamefont
  {Capaz}}, \bibinfo {author} {\bibfnamefont {A.}~\bibnamefont {{Dal Pino}}}, \
  and\ \bibinfo {author} {\bibfnamefont {J.~D.}\ \bibnamefont {Joannopoulos}},\
  }\href {\doibase 10.1103/PhysRevB.50.7439} {\bibfield  {journal} {\bibinfo
  {journal} {Physical Review B}\ }\textbf {\bibinfo {volume} {50}},\ \bibinfo
  {pages} {7439} (\bibinfo {year} {1994})}\BibitemShut {NoStop}%
\bibitem [{\citenamefont {Binder}\ and\ \citenamefont
  {Pasquarello}(2014)}]{Binder2014}%
  \BibitemOpen
  \bibfield  {author} {\bibinfo {author} {\bibfnamefont {J.~F.}\ \bibnamefont
  {Binder}}\ and\ \bibinfo {author} {\bibfnamefont {A.}~\bibnamefont
  {Pasquarello}},\ }\href {\doibase 10.1103/physrevb.89.245306} {\bibfield
  {journal} {\bibinfo  {journal} {Physical Review B}\ }\textbf {\bibinfo
  {volume} {89}},\ \bibinfo {pages} {245306} (\bibinfo {year}
  {2014})}\BibitemShut {NoStop}%
\bibitem [{\citenamefont {Backlund}\ and\ \citenamefont
  {Estreicher}(2007)}]{Backlund2007}%
  \BibitemOpen
  \bibfield  {author} {\bibinfo {author} {\bibfnamefont {D.}~\bibnamefont
  {Backlund}}\ and\ \bibinfo {author} {\bibfnamefont {S.}~\bibnamefont
  {Estreicher}},\ }\href {\doibase 10.1016/j.physb.2007.08.137} {\bibfield
  {journal} {\bibinfo  {journal} {Physica B: Condensed Matter}\ }\textbf
  {\bibinfo {volume} {401-402}},\ \bibinfo {pages} {163} (\bibinfo {year}
  {2007})}\BibitemShut {NoStop}%
\bibitem [{\citenamefont {Murin}\ \emph {et~al.}(1998)\citenamefont {Murin},
  \citenamefont {Hallberg}, \citenamefont {Markevich},\ and\ \citenamefont
  {Lindstr\"om}}]{Murin1998}%
  \BibitemOpen
  \bibfield  {author} {\bibinfo {author} {\bibfnamefont {L.~I.}\ \bibnamefont
  {Murin}}, \bibinfo {author} {\bibfnamefont {T.}~\bibnamefont {Hallberg}},
  \bibinfo {author} {\bibfnamefont {V.~P.}\ \bibnamefont {Markevich}}, \ and\
  \bibinfo {author} {\bibfnamefont {J.~L.}\ \bibnamefont {Lindstr\"om}},\
  }\href {\doibase 10.1103/physrevlett.80.93} {\bibfield  {journal} {\bibinfo
  {journal} {Physical Review Letters}\ }\textbf {\bibinfo {volume} {80}},\
  \bibinfo {pages} {93} (\bibinfo {year} {1998})}\BibitemShut {NoStop}%
\bibitem [{\citenamefont {Svensson}\ and\ \citenamefont
  {Lindstr\"om}(1986{\natexlab{b}})}]{Svensson1986}%
  \BibitemOpen
  \bibfield  {author} {\bibinfo {author} {\bibfnamefont {B.~G.}\ \bibnamefont
  {Svensson}}\ and\ \bibinfo {author} {\bibfnamefont {J.~L.}\ \bibnamefont
  {Lindstr\"om}},\ }\href {\doibase 10.1103/physrevb.34.8709} {\bibfield
  {journal} {\bibinfo  {journal} {Physical Review B}\ }\textbf {\bibinfo
  {volume} {34}},\ \bibinfo {pages} {8709} (\bibinfo {year}
  {1986}{\natexlab{b}})}\BibitemShut {NoStop}%
\bibitem [{\citenamefont {Freysoldt}\ \emph {et~al.}(2009)\citenamefont
  {Freysoldt}, \citenamefont {Neugebauer},\ and\ \citenamefont
  {de~Walle}}]{Freysoldt2009}%
  \BibitemOpen
  \bibfield  {author} {\bibinfo {author} {\bibfnamefont {C.}~\bibnamefont
  {Freysoldt}}, \bibinfo {author} {\bibfnamefont {J.}~\bibnamefont
  {Neugebauer}}, \ and\ \bibinfo {author} {\bibfnamefont {C.~G.~V.}\
  \bibnamefont {de~Walle}},\ }\href {\doibase 10.1103/physrevlett.102.016402}
  {\bibfield  {journal} {\bibinfo  {journal} {Physical Review Letters}\
  }\textbf {\bibinfo {volume} {102}},\ \bibinfo {pages} {016402} (\bibinfo
  {year} {2009})}\BibitemShut {NoStop}%
\bibitem [{\citenamefont {Coutinho}\ \emph
  {et~al.}(2001{\natexlab{b}})\citenamefont {Coutinho}, \citenamefont {Jones},
  \citenamefont {Murin}, \citenamefont {Markevich}, \citenamefont
  {Lindstr\"om}, \citenamefont {\"Oberg},\ and\ \citenamefont
  {Briddon}}]{Coutinho2001a}%
  \BibitemOpen
  \bibfield  {author} {\bibinfo {author} {\bibfnamefont {J.}~\bibnamefont
  {Coutinho}}, \bibinfo {author} {\bibfnamefont {R.}~\bibnamefont {Jones}},
  \bibinfo {author} {\bibfnamefont {L.~I.}\ \bibnamefont {Murin}}, \bibinfo
  {author} {\bibfnamefont {V.~P.}\ \bibnamefont {Markevich}}, \bibinfo {author}
  {\bibfnamefont {J.~L.}\ \bibnamefont {Lindstr\"om}}, \bibinfo {author}
  {\bibfnamefont {S.}~\bibnamefont {\"Oberg}}, \ and\ \bibinfo {author}
  {\bibfnamefont {P.~R.}\ \bibnamefont {Briddon}},\ }\href {\doibase
  10.1103/physrevlett.87.235501} {\bibfield  {journal} {\bibinfo  {journal}
  {Physical Review Letters}\ }\textbf {\bibinfo {volume} {87}},\ \bibinfo
  {pages} {235501} (\bibinfo {year} {2001}{\natexlab{b}})}\BibitemShut
  {NoStop}%
\end{thebibliography}
%

\end{document}